\documentclass[prd,showpacs,preprintnumbers,eqsecnum,amsmath,amssymb,nofootinbib]{revtex4}
\usepackage{graphicx}% Include figure files
\usepackage{bm}% bold math
\usepackage{epsfig}

\begin{document}
%\vspace{-15cm}
%\preprint{Preprint HD-THEP-03-xx.}
\title{
Masses and couplings of vector mesons from the pion electromagnetic, \\
weak, and $\pi\gamma$ transition form factors} 
\author{D. Melikhov$^{a}$\footnote{e-mail:~melikhov@thphys.uni-heidelberg.de}, 
O. Nachtmann$^a$\footnote{e-mail:~o.nachtmann@thphys.uni-heidelberg.de}, 
V. Nikonov$^b$\footnote{e-mail:~nikonov@thd.pnpi.spb.ru}, 
and 
T. Paulus$^a$\footnote{Now at Philips, e-mail:~t.paulus@philips.com}}
\affiliation{
$^a$ Institut f\"ur Theoretische Physik, Universit\"at Heidelberg, 
Philosophenweg 16,  D-69120, Heidelberg, Germany\\
%$^b$ Nuclear Physics Institute, Moscow State University, 119991, Moscow, Russia\\
$^b$ Petersburg Nuclear Physics Institute, Gatchina, 188300 St.-Petersburg, Russia}
\date{\today}
\begin{abstract}
We analyse the pion electromagnetic, charged-current, and $\pi\gamma$ transition 
form factors at timelike momentum transfers $q$, $q^2=s\le 1.4$ GeV$^2$, 
using a dispersion approach. 
We discuss in detail the propagator matrix of the photon-vector-meson system 
and define certain reduced amplitudes, or vertex functions, describing the coupling of
this system to final states.  
We then apply the derived analytic expressions to the analysis of the recent 
$e^+e^-\to \pi^+\pi^-$, $\tau^-\to \pi^-\pi^0\nu_\tau$, 
and $e^+e^-\to \pi^0\gamma$ data. 
We find the reduced amplitudes for the coupling of the photon and vector mesons to two
pseudoscalars to be constant, independent of $s$, in the range considered, indicating a
"freezing" of the amplitudes for $s\le 1$ GeV. 
The fit to the form factor data leads to the following values of 
the Breit-Wigner resonance masses 
$m_{\rho^-}=775.3\pm 0.8$ MeV, 
$m_{\rho^0}=773.7\pm 0.6$ MeV and 
$m_\omega=782.43\pm 0.05$ MeV, 
where the errors are only statistical. 
%We obtain also pole masses and widths of the $\rho$ and $\omega$. 
\end{abstract}
\maketitle

%%%\newpage
\section{Introduction}
The pion elastic and transition form factors at timelike momentum
transfers provide an important source of information about the masses and coupling 
constants of vector meson resonances $\rho$, $\omega$, $\phi$, etc. 
However a reliable extraction of the resonance parameters from the experimental data is 
a complicated problem. The reason is that the direct QCD-based calculation of the 
form factors in terms of the resonance parameters is not possible yet, and hence one 
has to use approximate approaches. A typical procedure of extracting the vector 
meson parameters is as follows: One relies on some theoretical formula for the form 
factor in terms of the vector-meson masses and couplings, and tries to extract their 
numerical values by fitting the experimental data. 
Most of the approaches providing theoretical inputs for the form factors in the region of 
vector meson resonances 0.7 to 1.2 GeV fall into the two big classes: approaches based on 
vector meson dominance picture \cite{vmd,klz,gs,kuhn,weise,vmd-mod}
and approaches based on inclusion of  
vector mesons into the Chiral perturbation theory framework \cite{chpt,piff,pich1}. 

No need to say that the extracted values of the meson parameters depend on the 
theoretical models used in this procedure; moreover, approximate formulas for the form 
factors in terms of the meson parameters inevitably introduce a 
systematic error which is extremely hard to control. 

Neglecting systematic uncertainties may lead to controversies in the determination of 
the resonance parameters. To illustrate this statement, let us turn to 
Table \ref{table:pdg} which presents the vector meson masses as quoted 
in the last three editions of the Particle Data Group \cite{pdg-old,pdg}. 
\begin{table}[th]
\caption{\label{table:pdg}
Values of the meson masses as quoted in last three editions of PDG \cite{pdg-old,pdg}.}
\centering
\vspace{.5cm}
\begin{tabular}{|c|c|c|c|}
\hline
         &  1998 & 2000 & 2002 \\
\hline
$\rho^0$ &  770.0 $\pm$0.8   &  769.3 $\pm$0.8   &  771.1 $\pm$0.9  \\
$\omega$ &  781.94$\pm$0.12  &  782.57$\pm$0.12  &  782.57$\pm$0.12   \\  
\hline
\end{tabular}
\end{table}
One clearly sees a very small error 
and relatively sizeable 'time-variations' of the average values such that some of 
the results from different editions are only marginally compatible with each other within 3$\sigma$. 
Moreover, the value of the $\rho$-meson mass $m_\rho=775.9\pm 0.5$ MeV as extracted from 
$\tau$-decays and $e^+e^-$ annihilation data is well above the value of the $\rho$-meson mass 
as obtained by averaging all data \cite{pdg}. 

The most natural explanation of this puzzle is that the systematic errors due to reliance on 
theoretical models may be underestimated. 
It is clear that a reliable extraction of the resonance parameters may only be reached if as broad as 
possible a set of data and reactions is used for the analysis. Comparison with the experiment serves 
as a test and to some extent a justification of the theoretical models and approximations employed. 

The data on the $e^+e^-\to\pi^+\pi^-$, $\tau^-\to\pi^-\pi^0\nu_\tau$, 
and $e^+e^-\to\pi^0\gamma$ reactions 
open a good possibility for extracting the $\rho$ and $\omega$ masses 
and coupling constants and testing various theoretical models.

As a first step we discuss our dispersion-theoretical framework where we start from 
the photon $\gamma$ and the vector meson $V$ ($V=\rho,\omega,\ldots)$ fields and identify the relevant
amplitudes and couplings in a model-independent way. Then we introduce our model which 
takes into account only the most important contributions to the absorptive parts in the dispersion 
relations. In essence, (i) we keep only the contributions of the $\pi\pi$ and $KK$ intermediate states 
and (ii) assume certain reduced amplitudes to be independent of the c.m. energy in the resonance region. 
Our approach does not use any specific effective Lagrangian or other approximation scheme like the $1/N_c$
expansion. However, rigorous theoretical results, 
which are known or may become available in the future, may be easily implemented in our 
framework as improved representations for the reduced amplitudes. 

In the second step we apply our formalism to perform a simultaneous
analysis of the pion electromagnetic, charged-current, and $\pi\gamma$ form factors.
We obtain analytic expressions for the form factors in terms of the
resonance parameters, paying special attention to the existing ambiguities in their definitions. 
We then apply our results to the $F_\pi$, $F_\pi^+$, and $F_{\gamma\pi}$ data 
and extract masses and couplings of the vector mesons. 
In this paper we consider only the most recent data \cite{cleo,data1,data0,data2} 
for these form factors in the range $2m_\pi\le \sqrt{s}\le 1.2$ GeV. 
We shall demonstrate that these data are well described by our model allowing for 
a reliable extraction of the vector-meson parameters. A systematic analysis of all available 
form factor data using our model is left for a future work. Then also relative normalisation
uncertainties between various data sets will have to be considered.

The paper is organised as follows: Section 2 gives definitions and summarises important 
rigorous results for $F_\pi$ and $F_{\gamma\pi}$. 
A general treatment of the vector meson-photon mixing in the framework of the dispersion approach is 
given in Section III. Our model is formulated in Section IV. In Section V numerical results 
are presented. Conclusions are given in Section VI. Appendices contain the necessary technical details. 

%%%\newpage
\section{The pion electromagnetic, weak, and $\pi\gamma$ form factors}
\subsection{The electromagnetic  form factor}
The pion  electromagnetic  form factor is defined by 
\begin{eqnarray}
\label{em}
\langle \pi^+(p')|J_\mu(0)|\pi^+(p)\rangle=e(p'+p)_\mu F_\pi(q^2),\qquad q=p'-p, 
\end{eqnarray}
for $q^2<0$ and by 
\begin{eqnarray}
\label{2.2}
\langle \pi^+(p') \pi^-(p) |J_\mu(0)|0\rangle=e(p'-p)_\mu F_\pi(q^2),\qquad q=p'+p, 
\end{eqnarray}
for $q^2>0$. Here $J_\mu$ is the electromagnetic current and $e=\sqrt{4\pi\alpha_{e.m.}}$. 
The form factor is normalised as $F_\pi(0)=1$. 

As function of the complex variable $s=q^2$, 
the form factor $F_\pi(s)$ has a cut in the complex $s$-plane starting at the two-pion 
threshold $s=4 m_\pi^2$ which corresponds to two-pion intermediate states. 
There are also cuts related to $K\bar K$ intermediate states and multi-meson states ($3\pi$, etc). 
The form factor in the timelike region $(s>0)$ is  
\begin{eqnarray}
F_\pi(s+i\epsilon )=|F_\pi(s)|e^{i\delta_\pi(s)}, 
\end{eqnarray}
where $\delta_\pi(s)$ is the phase. 
For the theoretical description of the form factor in different regions of momentum transfers
different theoretical approaches are used. 

At large spacelike momentum transfers, $-q^2\to\infty$, perturbative QCD (pQCD) 
gives rigorous predictions for the asymptotic behaviour of the form factor  
\cite{bl}
\begin{eqnarray}
\label{pqcd}
F_\pi(q^2)\sim\frac{8\pi f^2_\pi\alpha_s(-q^2)}{-q^2}, 
\end{eqnarray}
where $\alpha_s$ is the QCD coupling parameter and $f_\pi=130.7\pm 0.4$ MeV \cite{pdg} 
is the pion decay constant defined by the relation 
\begin{eqnarray}
\langle 0|\bar d(0)\gamma_\mu\gamma_5 u(0)|\pi^+(p)\rangle=ip_\mu f_\pi.
\end{eqnarray}
As the spacelike momentum transfer becomes smaller, 
the form factor turns out to be the result of the interplay of perturbative
and nonperturbative QCD effects, 
with a strong evidence that nonperturbative QCD effects dominate 
in the region $0\le -q^2\le 10$ GeV$^2$ \cite{amn,pigamma}. The picture based on the concept 
of constituent quarks which effectively account for nonperturbative dynamics 
has proven to be efficient for the description of the form factor in
this region (see for instance \cite{qm}). 

At large timelike momentum transfers, $s\ge 10\div 20$ GeV$^2$, $F_\pi(s)$ can be obtained from the 
analytic continuation of the pQCD formula (\ref{pqcd}). 
At small timelike momentum transfers the situation is more complicated 
since there dynamical details of the confinement mechanism are crucial. 
Quarks and gluons are no longer the degrees of freedom of QCD leading to a 
simple description of the form factor. 
At timelike momentum transfers we are essentially in the region of hadronic singularities 
and typically one relies on methods based on hadronic degrees of freedom. 
In the region of interest to us here, $q^2=0\div 1.5$ GeV$^2$, the 
lightest pseudoscalar mesons are most important. There are many approaches to understand 
the behaviour of the pion form factor at these timelike momentum transfers.

A popular approach is based on the vector meson dominance (VMD) model \cite{vmd}. 
In the simplest version one considers just the contribution of the $\rho$-meson pole, 
which in combination with the normalisation condition $F_{\pi}(s=0)=1$ leads to  
\begin{eqnarray}
\label{piff_a}
F_\pi(s)=\frac{m_\rho^2}{m_\rho^2-s}. 
\end{eqnarray}
This simple formula works with a reasonable accuracy both for small spacelike momentum transfers 
and timelike momentum transfers below the $\pi\pi$ threshold: 
$-1$ GeV$^2\le s\le 4m_\pi^2$.  
For $s$ near the $\pi\pi$ threshold one should take into account effects of the 
virtual pions. In this region, the momenta of the intermediate pions are small and a 
consistent description of the form factor is provided by chiral perturbation theory 
(ChPT) \cite{chpt}, the effective theory for QCD at low energies.

For higher $s$, in the region of $\rho$ and $\omega$ resonances,  
a similar rigorous treatment of the form factor is still lacking, 
and one has to rely on model considerations. 
Gounaris and Sakurai (GS) \cite{gs} obtained the expression for the 
$\rho$-meson contribution to the pion form factor which takes into account 
the $\rho$-meson finite width due to the virtual pions. The Gounaris-Sakurai form factor 
may be written in the form 
\begin{eqnarray}
\label{piff_b}
F_\pi(s)=\frac{m_\rho^2-B^{\rm GS}_{\rho\rho}(0)}{m_\rho^2-s-B^{\rm GS}_{\rho\rho}(s)}. 
\end{eqnarray}
The function $B^{\rm GS}_{\rho\rho}(s)$ corresponds to the two-pion loop diagram, but one can 
easily add the $K\bar K$ loop too, see Appendix \ref{app:B}. 
The corresponding Feynman integral is linearly divergent, but its imaginary part is 
defined in a unique way. The real part may then be reconstructed by a doubly-subtracted 
dispersion representation. The Gounaris-Sakurai formula corresponds to the following prescription 
of fixing the subtraction constants  
\begin{eqnarray}
\label{piff_gsb}
{\rm Re}\; B^{\rm GS}_{\rho\rho}(s)|_{s=m_\rho^2}=0,\qquad 
\frac{d}{ds}{\rm Re}\; B^{\rm GS}_{\rho\rho}(s)|_{s=m_\rho^2}=0. 
\end{eqnarray}
The phase of the GS form factor 
\begin{eqnarray}
\label{phase}
{\rm tan}\; \delta (s)=\frac{{\rm Im}B^{\rm GS}_{\rho\rho}(s)}{m_\rho^2-s-{\rm Re} B^{\rm GS}_{\rho\rho}(s)}. 
\end{eqnarray}
agrees well with the experimental data in the region $4m_\pi^2<s<0.9$ GeV$^2$. 

The Gounaris-Sakurai form factor (\ref{piff_b}) satisfies the relation $F_\pi(0)=1$. But it turns 
out that it does not have enough flexibility to give at the same time a good description at the 
peak of the $\rho$ resonance, see Appendix \ref{app:B}.

Near the $\rho$-meson peak the $\rho$-meson contribution to the pion form factor 
can be expressed in terms of the $\gamma\to\rho\to\pi\pi$ matrix element as follows: 
\begin{eqnarray}
\label{piff_c}
F_\pi(s)=\frac{\frac{1}{2}g_{\rho\to\pi\pi}f_\rho\;m_\rho}{m_\rho^2-s-B^{\rm GS}_{\rho\rho}(s)}. 
\end{eqnarray}
Here $g_{\rho\pi\pi}$ and $f_\rho$ are defined according to 
\begin{eqnarray}
\langle \pi^+(p')\pi^-(p)|T|\rho(q,\varepsilon)\rangle=
-\frac{1}{2}g_{\rho\to \pi\pi} \;\varepsilon_\mu\cdot (p'-p)^\mu,   
\qquad%\\
\label{fV}
\langle 0|J_\mu(0)|\rho^0(q,\varepsilon)\rangle=ef_\rho m_\rho \varepsilon_\mu, 
\end{eqnarray}
where $\varepsilon_\mu$ is the $\rho$-meson polarization 
and $q$ is the 4-momentum vector. 
Now $|F_\pi(s)|$ from (\ref{piff_c}) describes well the data 
for $s\simeq m_\rho^2$. But extrapolating the Eq. (\ref{piff_c}) to $s=0$ 
violates the normalisation condition $F_\pi(0)=1$ that is unacceptable. 

Note that the  definitions of $g_{\rho\to\pi\pi}$ and $f_\rho$ in (\ref{fV}) are not really 
appropriate since they are based on $\rho$-meson states. 
But the $\rho$ meson is a resonance and has no asymptotic states. Our precise definitions 
of $g_{\rho\to\pi\pi}$ and $f_\rho$ will be given below in Section IV. 

Thus, neither (\ref{piff_b}) nor (\ref{piff_c}) are suitable for the analysis of 
the form factor data for all $s=0\div 1.5$ GeV$^2$. 
There were many attempts to modify the vector meson dominance or to use related approaches in order to 
bring the results on the pion form factor in agreement with the data
(see \cite{weise,vmd-mod,pich1} and papers quoted therein). The pion form factor in the region 
$s=0\div 1.5$ GeV$^2$ is one of the main sources for obtaining masses and coupling constants of 
vector mesons. 
However, with different assumptions on the form of the vector-resonance contribution to 
the pion form factor one obtains different values of masses and couplings. 
Therefore a consistent description of the pion form factor in this region in terms of the low-lying
mesons ($\pi, K, \rho, \omega$) is crucial for extracting reliable values of these parameters. 
Interesting results have been obtained by the authors of \cite{weise} who noticed that 
an effective momentum-dependent $\rho\gamma$ coupling appears in the framework of the effective 
Lagrangian approach. This momentum-dependent $\rho\gamma$ coupling also  
improves the description of the pion form factor at timelike momentum transfers in the region 
$0< q^2 <1$ GeV$^2$. 

%A powerful tool for the theoretical analysis are dispersion representations.  
%Assuming a dispersion relation in $q^2$ with one subtraction, we obtain 
%\begin{eqnarray}
%\label{ffdispsub}
%F_{\pi}(q^2)=F_{\pi}(0)+q^2 \int \frac{ds}{\pi} 
%\frac{{\rm Im}\; F_{\pi}(s)}{s(s-q^2)}. 
%\end{eqnarray}
%Neglecting the vector-meson mixing and finite-width effects, one finds 
%for the $\rho$ contribution to the form factor 
%\begin{eqnarray}
%\label{piff_c}
%F_\pi(q^2)=1+\frac{1}{2}g_{\rho\to\pi\pi}\frac{q^2f_\rho}{m_\rho}
%\frac{1}{m_\rho^2-q^2}. 
%\end{eqnarray}
%Here $g_{\rho\pi\pi}$ and $f_\rho$ are defined according to  
%\begin{eqnarray}
%\langle \pi(k_1)\pi(k_2)|T|\rho(\varepsilon,k)\rangle&=&
%\frac{1}{2}g_{\rho\to \pi\pi} \;\varepsilon_\mu\cdot (k_1-k_2)^\mu,   
%\\
%\langle 0|J_\mu|\rho^0(\varepsilon,k)\rangle&=&f_\rho m_\rho \varepsilon_\mu, 
%\end{eqnarray}
%where $\varepsilon_\mu$ is the $\rho$-meson polarization vector 
%and $k$ is the 4-momentum vector. 

Clearly, to achieve a realistic description of the form factor, one has to account for 
vector meson finite-width and mixing effects. 
This may be done in a consistent way within a dispersion approach and will be the subject 
of Section III. 

\subsection{The weak form factor}
The $\pi^-\to \pi^0$ weak transition form factors parametrise 
the charged-current transition amplitude as follows 
\begin{eqnarray}
\label{weak}
\langle \pi^+(p)\pi^0(p')|\bar u(0)\gamma_\mu d(0)|0\rangle=
{\sqrt2}F_\pi^+(q^2) 
(p'-p)_\mu + {\sqrt2}F_\pi^-(q^2)q_\mu.
\end{eqnarray}
In the limit of the exact isospin symmetry we have $F_\pi^-=0$ and 
$F_\pi^+=F_\pi|_{\rm isovector}$. In practice we expect that the electromagnetic form factor 
$F_{\pi}$ 
%(which contains isovector and isoscalar parts) 
should be close to $F_\pi^+$ for 
$0\le q^2\le 1$ GeV$^2$ except for the region of the $\omega$ resonance. 
The form factor $F_\pi$ contains an important isospin-violating contribution of the  
$\omega$ resonance, whereas there is no contribution analogous to $\omega$ in $F_\pi^+$. 

%****************************************************************
\subsection{The $\pi\gamma$ form factor}
We shall be interested in the process $e^+e^-\to\gamma^*\to \pi^0\gamma$ 
where one of the photons is real and the other is virtual.  
The form factor $F_{\gamma\pi}$ relevant for this process is defined 
\cite{pigamma,pigamma2} according to 
\begin{eqnarray}
\label{2.13}
\langle \pi^0(p)\gamma(q',\varepsilon)|J_{\mu}(0)|0\rangle=
e^2 \epsilon_{\alpha\beta\mu\nu}\varepsilon^{*\nu}q^\alpha q'^{\beta}
F_{\gamma\pi}(q^2). 
\end{eqnarray}

In terms of this form factor the $e^+e^-\to\pi^0\gamma$ cross section reads 
\begin{eqnarray}
\label{cross-section}
\sigma_{e^+e^-\to\pi^0\gamma}(q^2)
=\frac{2}{3}\pi^2\alpha_{e.m.}^3\left(1-\frac{m_\pi^2}{q^2}\right)^3|F_{\gamma\pi}(q^2)|^2.
\end{eqnarray}
In the chiral limit, $m_\pi^2=0$, the value of the form factor for $q^2=0$ 
is fixed by the Adler-Bell-Jackiw anomaly \cite{abj}  
\begin{eqnarray}
\label{fpigamma0}
F_{\gamma\pi}(0)|_{m_\pi^2=0}=\frac{1}{2\sqrt{2}\pi^2f_\pi}. 
\end{eqnarray}
In reality the pion is not massless, but still the anomaly provides a very good 
description of the observed $\pi^0\to\gamma\gamma$ decay rate.

%  
%$F^{\rm exp}_{\gamma\gamma\pi}/F^{\rm anomaly}_{\gamma\gamma\pi}=
%(0.26\pm 0.07)/0.275$. 
%
We shall therefore use the value (\ref{fpigamma0}) also for the physical pion 
in order to fix the form factor $F_{\gamma\pi}(0)$ in our analysis.  

%***************
In the region of large spacelike momentum transfer $q$, the form factor can be calculated 
from pQCD with the result 
\begin{eqnarray}
F_{\gamma\pi}(q^2)\sim  \frac{\sqrt{2f_\pi}}{-q^2}. 
\end{eqnarray}
Brodsky and Lepage \cite{brodsky} proposed a simple formula 
\begin{eqnarray}
F_{\gamma\pi}(q^2)=\frac{\sqrt{2f_\pi}}{4\pi^2f_\pi^2-q^2}, 
\end{eqnarray}
which interpolates between $q^2=0$ and $q^2\to-\infty$ and works well for all 
$q^2<0$. This formula may be written as 
\begin{eqnarray}
F_{\gamma\pi}(q^2)=\frac{\sqrt{2f_\pi}}{M_{\rm Res}^2-q^2}, 
\end{eqnarray}
with $M_{\rm Res}=2\pi f_\pi=880$ MeV, not too far from the masses of $\rho$ and $\omega$ 
which give the dominant resonance contribution to the form factor. 

To describe the form factor in the region $0<q^2\le 1.5$ GeV$^2$, we should  
again use the meson degrees of freedom. For a realistic description of the 
form factors we must take into account finite-width and meson mixing effects. 

%%%\newpage
\section{Mixing of vector mesons: Propagator matrix and vertex functions}
In this section we present model-independent considerations on the mixing of the photon with 
vector mesons. 

Since vector mesons are unstable particles, one of the possibilities is to 
start with hypothetical stable states, which then get a width by inclusion 
of some interactions. This is an inherently perturbative picture which emerges for instance 
when the $1/N_c$ expansion is used. 

We shall avoid such a perturbative approach and instead start with properly defined renormalised 
field operators with the quantum numbers of the vector mesons we are interested in. Clearly, such 
field operators can be defined in the framework of QCD. As the second step, we shall analyse the propagator matrix 
describing the mixing of these vector-meson fields with the photon field. Then, we define 
certain transition amplitudes (or vertex functions) which are one-particle irreducible in
the $s$ channel, 
and establish the connection between these vertex functions and the experimentally measured 
form factors. These considerations are fully general and do not include any model assumptions. 
As the next step our model is formulated making certain assumptions for these vertex functions. 
This procedure is similar to the one used in \cite{nachtmann} in the discussion of the 
$\gamma$-$J/\psi$ mixing. 

Let us consider the photon field $A_\mu(x)$ and a set of hermitian neutral vector-meson fields $V_\mu^{(j)}(x)$, 
($j=2,...,n$). For convenience of notation we set $V_\mu^{(1)}(x)=A_\mu(x)$. 

%In our applications we shall 
%mainly consider the case $n=3$ with $V^{(1)},V^{(2)},V^{(3)}$ the $\gamma,\rho$ and $\omega$ fields. 

The fields $V_\mu^{(j)}(x)$ ($j=1,...,n$) have the same quantum numbers and therefore will have a 
$n\times n$ propagator matrix describing their mixing
\begin{eqnarray}
\label{3.1}
\Delta_{\mu\nu}^{(j,k)}(q)=\frac{1}{i}\int d^4x\, e^{iqx}\langle 0|T^*\{V_\mu^{(j)}(x)\,V_\nu^{(k)}(0)\}|0\rangle. 
\end{eqnarray}
Here $T^*$ is the covariant version of the $T$ product, 
see for instance \cite{weinberg}. Using a covariant gauge for the photon, we can separate 
$\Delta_{\mu\nu}^{(j,k)}$ into transverse and longitudinal parts as follows 
\begin{eqnarray}
\label{3.2}
\Delta_{\mu\nu}^{(j,k)}=\left(-g_{\mu\nu}+\frac{q_\mu q_\nu}{q^2+i\epsilon }\right)\Delta_{T}^{(j,k)}(q^2)
-\frac{q_\mu q_\nu}{q^2+i\epsilon }\ \Delta_{L}^{(j,k)}(q^2). 
\end{eqnarray}
The matrices 
\begin{eqnarray}
\Delta_{T,L}(q^2)=\left(\Delta_{T,L}^{(j,k)}(q^2)  \right)
\end{eqnarray}
are analytic in the complex $q^2$-plane with cuts on the positive real axis.  
We shall always work to leading order in the electromagnetic interaction. 
Then the leftmost cut starts at $q^2=4m_\pi^2$, the two-pion threshold. 
Similarly, the transverse and the longitudinal structures
can be isolated in the inverse propagator matrix
\begin{eqnarray}
\label{5.5a}
\left(\Delta^{-1}(q)\right)_{\mu\nu}^{(j,k)}=
\left(-g_{\mu\nu}+\frac{q_\mu q_\nu}{q^2+i\epsilon }\right)\left(\Delta^{-1}_{T}(q^2)\right)^{(j,k)}
-\frac{q_\mu q_\nu}{q^2+i\epsilon }\ \left(\Delta^{-1}_{L}(q^2)\right)^{(j,k)}. 
\end{eqnarray}
The propagator matrix satisfies several general relations. 
\begin{itemize}
\item
Translation invariance of the vacuum gives 
\begin{eqnarray}
\label{5.6} 
\Delta_{\mu\nu}^{(j,k)}(q)=\Delta_{\nu\mu}^{(k,j)}(-q). 
\end{eqnarray}
\item
CPT invariance gives
\begin{eqnarray}
\label{ii}
\Delta_{\mu\nu}^{(j,k)}(q)=\Delta_{\mu\nu}^{(j,k)}(-q). 
\end{eqnarray}
\item
T-invariance of strong and electromagnetic interactions gives
\begin{eqnarray}
\label{iii}
\Delta_{\mu\nu}^{(j,k)}(q^0,\vec q)=\Delta^{(j,k){\mu\nu}}(-q^0,\vec q). 
\end{eqnarray}
\end{itemize}
From (\ref{5.6}) we find that the matrices $\Delta_{T,L}$ must be symmetric
\begin{eqnarray}
\left(\Delta_{T,L}(q^2)\right)^T=\Delta_{T,L}(q^2),  
\end{eqnarray}
whereas (\ref{ii}) and (\ref{iii}) are satisfied automatically and give no restrictions. 

Using next the hermiticity of the fields $V^{(j)}_\mu(x)$ we get the unitarity relation for the 
propagator matrix 
\begin{eqnarray}
\label{5.10}
\Delta_{\mu\nu}^{(j,k)}(q)-\left(\Delta_{\nu\mu}^{(k,j)}(q)\right)^*
&=&
-i\int d^4x\, e^{iqx}
\langle 0|\left\{V^{(j)}_\mu(x) V^{(k)}_\nu(0) + V^{(k)}_\nu(0) V^{(j)}_\mu(x)\right\}|0\rangle 
\nonumber\\
&=&-i\sum\limits_X
\left\{
(2\pi)^4\delta^{(4)}(q-p_X) 
\langle 0|V^{(j)}_\mu(0)|X(p_X)\rangle \langle X(p_X)|V^{(k)}_\nu(0)|0\rangle\right.
\nonumber\\
&&\qquad\quad+\left.(2\pi)^4\delta^{(4)}(q+p_X) 
\langle 0|V^{(k)}_\nu(0)|X(p_X)\rangle \langle X(p_X)|V^{(j)}_\mu(0)|0\rangle
\right\}
\end{eqnarray}
Here we have inserted a complete set of asymptotic (in strong interactions) states $|X(p_X)\rangle$, where $p_X$ is the 
four-momentum. Note that the states $|X(p_X)\rangle$ contain pions and kaons, but no $\rho$ or $\omega$
mesons since the latter are unstable and thus have no asymptotic states. 

Let us now define for all states $|X(p_X)\rangle$ the reduced, or amputated, matrix elements 
$\langle X(p_X)||V^{(j)}_\mu||0\rangle$ by taking out of 
$\langle X(p_X)|V^{(j)}_\mu|0\rangle$ all $s$ channel $V$-propagator terms: 
\begin{eqnarray}
\label{5.11}
\langle X(p_X)||V^{(j)}_\mu||0\rangle=\langle X(p_X)|V^{(i)\nu}|0\rangle
\left(\Delta^{-1}(p_X)\right)^{(i,j)}_{\nu\mu}. 
\end{eqnarray}
Here and in the following we use the summation convention. The inverse of (\ref{5.11}) reads 
\begin{eqnarray}
\label{5.12}
%(5.12)\qquad
\langle X(p_X)|V^{(j)}_\mu|0\rangle=\langle X(p_X)||V^{(i)\nu}||0\rangle
\Delta^{(i,j)}_{\nu\mu}(p_X). 
\end{eqnarray}
The reduced matrix elements, or vertex functions, $\langle X(p_X)||(V^{(j)}_\mu||0\rangle$  
are one-$V$ irreducible in the s-channel. 

It is convenient to define the transverse and the longitudinal components of the vertex functions 
\begin{eqnarray}
\label{5.13}
\langle X(p_X)||V^{(j)}_{T\mu}||0\rangle&=&
\langle X(p_X)||V^{(j)\nu}||0\rangle\left(g_{\nu\mu}-\frac{(p_X)_\nu (p_X)_\mu}{(p_X)^2}\right), 
\nonumber\\
\langle X(p_X)||V^{(j)}_{L}||0\rangle&=&
\langle X(p_X)||V^{(j)\nu}||0\rangle\frac{(p_X)_\nu}{\sqrt{(p_X)^2}}. 
\end{eqnarray}
Now insert (\ref{5.12}) into the unitarity relation (\ref{5.10}). Considering first 
(\ref{5.10}) for $q^0>0$, we obtain 
\begin{eqnarray}
%\label{5.15}
\Delta_{\mu\nu}^{(j,k)}(q)-\left(\Delta_{\nu\mu}^{(k,j)}(q)\right)^*
=
-i\sum\limits_X(2\pi)^4\delta^{(4)}(q-p_X)
\left(\Delta_{\mu'\mu}^{(j',j)}\right)^*
\langle X(p_X)||V^{(j')\mu'}||0\rangle^*
\langle X(p_X)||V^{(k')\nu'}||0\rangle
\Delta_{\nu'\nu}^{(k',k)}.\nonumber\\
\label{5.15} 
\end{eqnarray}
Multiplying (\ref{5.15}) by $\left(\Delta^{-1}\right)^\dagger$ from the left and 
by $\Delta^{-1}$ from the right we come to the unitarity relation for the inverse propagator 
(\ref{5.5a}) as 
\begin{eqnarray}
\label{5.16}
\frac{1}{2i}\left\{\Delta^{-1}_{T,L}(q^2)-\left(\Delta^{-1}_{T,L}(q^2)\right)^\dagger\right\}
=D_{T,L}(q^2), 
\end{eqnarray}
where the discontinuity matrices $D_{T,L}(q^2)$ are given by 
\begin{eqnarray}
\label{5.17}
D_{T}^{(j,k)}(q^2)&=&
-\frac{1}{6} 
\sum\limits_X(2\pi)^4\delta^{(4)}(q-p_X)
\langle X(p_X)||V^{(j)}_{T\lambda}||0\rangle^*
\langle X(p_X)||V^{(k)\lambda}_T||0\rangle\\
\label{5.18}
D_{L}^{(j,k)}(q^2)&=&
-\frac{1}{2} 
\sum\limits_X(2\pi)^4\delta^{(4)}(q-p_X)
\langle X(p_X)||V^{(j)}_L||0\rangle^*
\langle X(p_X)||V^{(k)}_L||0\rangle.
\end{eqnarray}
The discontinuity matrices satisfy the relations 
\begin{eqnarray}
\label{5.19}
D_{T}(q^2)=\left(D_{T}(q^2)\right)^T&=&\left(D_{T}(q^2)\right)^\dagger,  
\nonumber\\
D_{T}(q^2)&=&0, \qquad {\rm for} \quad q^2<4m_\pi^2,  
\nonumber\\
D_{T}(q^2)&\ge &0, \qquad {\rm for} \quad q^2\ge 4m_\pi^2, 
\end{eqnarray}
and 
\begin{eqnarray}
\label{5.20}
D_{L}(q^2)=\left(D_{L}(q^2)\right)^T&=&\left(D_{L}(q^2)\right)^\dagger, 
\nonumber\\
D_{L}(q^2)&=&0, \qquad {\rm for} \quad q^2<4m_\pi^2,  
\nonumber\\
D_{L}(q^2)&\le &0, \qquad {\rm for} \quad q^2\ge 4m_\pi^2. 
\end{eqnarray}
Considering in (\ref{5.10}) the case $q^0 < 0$, inserting (\ref{5.12}) and using  (\ref{5.6}) 
we find exactly the same relations (\ref{5.16}) -- (\ref{5.20}). 

In our applications the longitudinal part $\Delta_L(q^2)$ plays no role, so we concentrate 
on the transverse part $\Delta_T(q^2)$ in the following. 

The analytic properties of $\Delta^{-1}_T(q^2)$ allow us to write a dispersion relation for it, 
which we assume to be convergent with two subtractions: 
\begin{eqnarray}
\label{5.21}
%(5.21)\qquad\qquad
\Delta^{-1}_{T}(q^2)=-M^2+K q^2+(q^2)^2\frac{1}{\pi}
\int\limits_{4m_\pi^2}^{\infty}ds\frac{D_T(s)}{s^2(s-q^2-i\epsilon )}. 
\end{eqnarray}
Here the subtraction terms $M^2$ and $K$ have to be constant real symmetric matrices 
\begin{eqnarray}
\label{5.22}
M^2=(M^2)^T=(M^2)^*, \qquad K=K^T=K^*. 
\end{eqnarray}
This is as far as we can come with a general analysis of the propagator matrix. 

In the next section we shall analyse the amplitude 
$\langle X(p_X)|A_\mu|0\rangle$ for the electromagnetic field and the 
state $|X(p_X) \rangle$ being the $\pi\pi$ state, which gives the pion form factor. 
The equation (\ref{5.12}) with $j=1$ represents this amplitude in terms of the vertex functions  
$\langle X(p_X)||V^{(i)}_\mu||0\rangle$ and the propagator matrix $\Delta_{\mu\nu}$ for which 
we have the dispersion representation. 
 
The merit of this representation is that different types of singularities are isolated in different 
quantities: the propagator matrix contains the resonance poles which lead to "fast" variations of the 
form factors in the resonance region; the reduced amplitudes are free from these singularities and
therefore represent slowly varying functions in the resonance region. 
To go further with the form factors we need some dynamical inputs for the vertex functions 
$\langle X(p_X)||V^{(j)\mu}||0\rangle$ and for the matrices $M^2$ and $K$, see Section IV.  

Before going to the details of the model, note that we are free to change the 
basis for the fields. Defining new fields 
\begin{eqnarray}
\label{5.23}
\tilde V^{(j)}_{\mu}(x)=C_{jk}\,V^{(k)}_{\mu}(x), 
\end{eqnarray}
with $C=(C_{jk})$ a real non-singular $n\times n$ matrix,\footnote{
In field theory we have also a freedom to make more complicated redefinitions of the fields, for
instance 
\begin{eqnarray}
%\label{5.26}
\nonumber
\tilde V_{\mu}(x)=(1+c\,\Box)V_{\mu}(x), \qquad c=const. 
\end{eqnarray}
Such transformations will change the $q^2$ behavior of the propagators and the $p_X^2$ behavior of the
vertex functions. In the present article we will not explore further the possibility of
such field redefinitions.} 
we get the propagator matrix of the new fields 
\begin{eqnarray}
\label{5.24}
\tilde \Delta_{\mu\nu}(q)=C \Delta_{\mu\nu}(q) C^T.
\end{eqnarray}
This leads to 
\begin{eqnarray}
\label{5.24a}
\tilde \Delta^{-1}_{T,L}(q^2)=\left(C^{-1}\right)^T \Delta^{-1}_{T,L}(q^2) C^{-1}, 
\end{eqnarray}
\begin{eqnarray}
\label{5.25}
%(5.25)\qquad
\langle X(p_X)||\tilde V^{(j)}_\mu||0\rangle=\langle X(p_X)|| V^{(k)}_\mu ||0\rangle C^{-1}_{kj}. 
\end{eqnarray}
The freedom of the field redefinition (\ref{5.23})  can and will be used to impose certain constraints 
on the matrices $M^2$ and $K$ in (\ref{5.21}). If $K$ is a positive-definite matrix - 
as it should be from the positivity of the metric for physical states in the Hilbert space - 
we can, for instance, diagonalise $K$ and $M^2$ simultaneously by a transformation (\ref{5.23}). 
The procedure to achieve this is completely analogous to the introduction of normal coordinates 
in the problem of small oscillations around a stable minimum of the potential in mechanics 
(see for instance \cite{goldstein}). 

We should, however, be careful with redefinitions of the photon field $V_\mu^{(1)}=A_\mu$. 
A redefined photon field containing components proportional to the strong interacion vector fields 
$V_\mu^{(j)}$ with $j>1$ will induce a direct quark-lepton coupling. 
We think this is unacceptable. 
The conditions which allow to avoid this and to guarantee the massless photon and the correct charge
normalisation are summarised in Appendix \ref{app:A} and lead to 
\begin{eqnarray}
\label{44a}
M_{1j}^2&=&0, \qquad j=1,...n, \\
\label{44b}
K_{11}&=&1. 
\end{eqnarray}

%%%\newpage 
\section{\label{sec4}The $\gamma-\rho-\omega$ system}
\subsection{The model}
We now apply the general considerations of the previous section to the system containing the
photon field and the vector meson $\rho$ and $\omega$ fields, 
\begin{eqnarray}
\label{6.1}
V^{(1)}_{\mu}(x)=A_\mu(x), \qquad 
V^{(2)}_{\mu}(x)=\rho_\mu(x), \qquad 
V^{(3)}_{\mu}(x)=\omega_\mu(x). 
\end{eqnarray}
We suppose the field $\rho_\mu$ to be purely isovector, and $\omega_\mu$ to be purely isoscalar. 
The electromagnetic coupling and isospin breaking from different up and down quark masses 
in QCD will introduce non-diagonal terms in the propagator matrix. In the following we will 
frequently use the indices $\gamma$, $\rho$, and $\omega$ instead of 1, 2, 3. 

The $3\times 3$ matrix $M^2$ of (\ref{5.21}), (\ref{5.22}) for our system has to satisfy 
(\ref{44a}). By a linear transformation (\ref{5.23}), but involving only the $\rho$ and
$\omega$ fields we can make $M^2$ diagonal.  

At this stage we define the $\rho$ and $\omega$ mass squared parameters 
$m_\rho^2$ and $m_\omega^2$ as zero points of the real parts of the diagonal terms of the inverse 
propagator matrix, that is by the relations  
\begin{eqnarray}
\label{6.2}
{\rm Re} \left(\Delta_T^{-1}\right)^{(\rho,\rho)}(m_\rho^2)=0,\qquad 
{\rm Re} \left(\Delta_T^{-1}\right)^{(\omega,\omega)}(m_\omega^2)=0.
\end{eqnarray} 
Then we choose the normalisation of the fields $\rho_\mu$ and $\omega_\mu$ such that 
the matrix $M^2$ has the form  
\begin{eqnarray}
\label{6.4}
{M}^2=
\left(\begin{array}{ccc} 
0 &  0                    &  0\\
0 & m_\rho^2              &  0\\
0 &  0            &  m_\omega^2
\end{array}\right).
\end{eqnarray} 

%As clear from the conditions (\ref{6.2}), the matrix $K$ depends on the dispersive part of the 
%inverse propagator matrix. So we shall not discuss $K$ separately, but rather add the 
%relevant subtraction terms linear in $s$ directly in $\Delta^{-1}$. 
%
%
%  MATRIX K
%
%\begin{eqnarray}
%\label{6.4a}
%{ K}=
%\left(\begin{array}{ccc} 
%1                   &  e\frac{f_\rho}{m_\rho}  &  e\frac{f_\omega}{m_\omega}\\
%e\frac{f_\rho}{m_\rho}       &  1                    & b_{\rho\omega}\\
%e\frac{f_\omega}{m_\omega}     & b_{\rho\omega}    &  1
%\end{array}\right). 
%\end{eqnarray} 
%We have introduced here the dimensionful parameters ${f_\rho}$ and ${f_\omega}$ which 
%correspond to the usually considered leptonic decay constants of the vector mesons. 
%Up to now the consideration was fully general.  
%
%The $\rho\omega$ mixing is essential only in the region of the $\rho$ and $\omega$, therefore 
%we can for practical applications set $\mu^2_{\rho\omega}=0$ in (\ref{6.4}). 
%This is our first assumption. The parameters $m_\rho^2$, $m_\omega^2$, 
%$f_\rho$, and $f_\omega$ will be the variational parameters of our analysis. 
%

To calculate the dispersive part of the inverse propagator, we must 
restrict the set of the intermediate states $|X(p_X)\rangle$ to be included in the 
unitarity relation (\ref{5.15}), and parametrise the reduced amplitudes of the fields $V^{(j)}_\mu$ 
between these states and the vacuum. 

{\bf Assumption 1}: as the intermediate states $|X(p_X)\rangle$ in the dispersion relation 
(\ref{5.21}) with $D_T$ given by (\ref{5.17}) we shall consider only $\pi^+\pi^-$, $3\pi$, 
$K^+ K^-$, and $K^0\bar K^0$ states.

For the $\pi^+\pi^-$ and $KK$ states we have   
\begin{eqnarray}
\label{6.5}
\langle \pi^+(k_1)\pi^-(k_2)||V_{T\mu}^{j}||0\rangle&=&g^{(j)}_{\pi\pi}(k_1-k_2)_\mu, 
\nonumber\\
\langle K^+(k_1)K^-(k_2)||V_{T\mu}^{j}||0\rangle&=&g^{(j)}_{KK}(k_1-k_2)_\mu,
\end{eqnarray}
where $g^{(j)}_{\pi\pi}$ and $g^{(j)}_{KK}$ are in general (slowly varying) functions of $(k_1+k_2)^2$.  

\noindent
{\bf Assumption 2}: in the region of interest we neglect the dependence of 
$g^{(j)}_{\pi\pi}$ and $g^{(j)}_{KK}$ on $(k_1+k_2)^2$ and assume all $g^{(i)}$ 
to be real constants 
\begin{eqnarray}
\label{6.6a} 
g^{(1)}_{\pi\pi}=e,\qquad 
g^{(2)}_{\pi\pi}=\frac{1}{2}g_{\rho\to\pi\pi}, \qquad 
g^{(3)}_{\pi\pi}=\frac{1}{2}g_{\omega\to\pi\pi}, 
\end{eqnarray}
and similarly for the $KK$ intermediate states 
\begin{eqnarray}
\label{6.6b}
&&g^{(1)}_{K^+K^-}=e,\qquad  g^{(1)}_{K^0\bar K^0}=0, \qquad 
%\nonumber\\&&
g^{(2)}_{K^+K^-}=-g^{(2)}_{K^0\bar K^0}=\frac{1}{2}g_{\rho\to KK}, \qquad 
%\nonumber\\&&
g^{(3)}_{K^+K^-}=g^{(3)}_{K^0\bar K^0}=\frac{1}{2}g_{\omega\to KK}. 
\end{eqnarray}
%After possible rescaling of the $\rho_\mu$ and $\omega_\mu$ fields 
Here $g^{(1)}_{\pi\pi}=g^{(1)}_{K^+K^-}=e$ and $g^{(1)}_{K^0\bar{K^0}}=0$ as required by
the charge normalisation of the $\pi^+$, $K^+$ and $K^0$, see Appendix \ref{app:C}. 

The decays $\rho\to K\bar K$ and $\omega\to K\bar K$ are forbidden kinematically at the
$\rho$ and $\omega$ peaks. This makes a direct determination of the corresponding coupling
constants $g_{\rho\to KK}$ and $g_{\omega\to KK}$ difficult. Therefore we use as additional
theoretical input the relations following from the approximate SU(3) flavour symmetry of
strong interactions and ideal mixing of the vector mesons 
\begin{eqnarray}
\label{su3}
g_{\rho\to KK}=g_{\omega\to KK}=\frac{1}{2}g_{\rho\to \pi\pi}.
\end{eqnarray}

In this paper we do not analyse the 3$\pi$ decays in detail. In the unitarity relation 
(\ref{5.15}), (\ref{5.16}) the 3$\pi$ intermediate states produce the width of the $\omega$, 
$\Gamma_{\omega}$, which is one of the fitting parameters. 

We now have to specify the matrix $K$ in (\ref{5.21}). 
The explicit form of the matrix $K$ is discussed in Appendix \ref{app:B}, and here we present the 
final form for the inverse propagator matrix in our model: 
\begin{eqnarray}
\label{6.7}
%(6.7)\nonumber\\
\Delta_T^{-1}(s)=
\left(\begin{array}{ccc} 
s                                               
&\qquad e\frac{f_\rho}{m_\rho}s+B_{\gamma\rho}(s)  
&\qquad  e\frac{f_\omega}{m_\omega}s+B_{\gamma\omega}(s)                     \\
e\frac{f_\rho}{m_\rho}s+B_{\gamma\rho}(s)       
&\qquad -m_\rho^2+s+B_{\rho\rho}(s)                
&\qquad s\,b_{\rho\omega}+B_{\rho\omega}(s)            \\
e\frac{f_\omega}{m_\omega}s+B_{\gamma\omega}(s) 
&\qquad s\,b_{\rho\omega}+B_{\rho\omega}(s)     
&\qquad  -m_\omega^2+s+B_{\omega\omega}(s)
\end{array}\right).
\end{eqnarray} 
The functions $B_{ij}$ are constructed by the doubly-subtracted dispersion integrals (\ref{5.21}) 
corresponding to the pion and kaon contributions and include also the relevant subtraction terms 
defined such that 
\begin{eqnarray}
\label{6.8} 
B_{ij}(s=0)=0, \qquad 
{\rm Re} B_{\rho\rho}(m_\rho^2)={\rm Re} B_{\gamma\rho}(m_\rho^2)={\rm Re} B_{\rho\omega}(m_\rho^2)=0, \qquad 
{\rm Re} B_{\omega\omega}(m_\omega^2)={\rm Re} B_{\gamma\omega}(m_\omega^2)=0. 
\end{eqnarray} 
For the functions $B_{ij}$ defined according to the conditions (\ref{6.8}), 
the dimensionful constants $f_\rho$  and $f_\omega$ correspond to our precise 
definitions of the leptonic decay constants of the vector mesons.
The detailed formulas for $B_{ij}$ are given in Appendix \ref{app:B}. 

%The $\rho\omega$ mixing effects 
%are essential only in the region of $q^2\simeq m_\rho^2,m_\omega^2$. 
%Because of that we cannot distinguish between the $\rho-\omega$ mixing described 
%by the constant term $\mu_{\rho\omega}^2$ and the $s$-dependent term 
%$s b_{\rho\omega}$. Therefore we shall keep only the term $s b_{\rho\omega}$ 
%and omit the term $\mu^2_{\rho\omega}$. 

The intermediate $\pi\gamma$ states do not contribute to the form factors to first order in the e.m. coupling. 
Nevertheless, we need the reduced $\pi\gamma$ amplitudes for the description of the $\pi\gamma$ transition form factor. 
The reduced $\pi\gamma$ amplitudes have the form 
\begin{eqnarray}
\label{6.5b}
\langle \pi(k_1)\gamma(k_2,\varepsilon)||V^{(j)}_{\mu}(0)||0\rangle=
e \epsilon_{\alpha\beta\mu\nu}\varepsilon^{*\nu}k_1^\alpha k_2^{\beta}
g^{(j)}_{\gamma\pi}, 
\end{eqnarray}
where $g^{(j)}_{\gamma\pi}$ in general depend on $(k_1+k_2)^2$. 
We assume the $g^{(j)}_{\gamma\pi}$ to be constant as we did for 
the $\pi\pi$ and $K\bar K$ couplings. 
Then $g^{(1)}_{\gamma\pi}$ is determined from the anomaly (see appendix C) 
\begin{eqnarray}
\label{4.11}
g^{(1)}_{\gamma\pi}=eF_{\gamma\pi}(0)=e\frac{1}{2\sqrt{2}\pi^2f_\pi}. 
\end{eqnarray}
The two additional dimensionless parameters $g_{\rho\to\gamma\pi}=m_\rho\,g^{(2)}_{\gamma\pi}$ 
and $g_{\omega\to\gamma\pi}=m_\omega\,g^{(3)}_{\gamma\pi}$ are assumed to be real. 

%The constants $f_\rho$ and $f_\omega$ appear as parameters describing one part of the matrix $K$ in (\ref{5.21}), 
%the other part of the matrix $K$ is related to the real part of the functions $B_{ij}$ responsible for fulfilling the relations (\ref{cond}). 
%For instance $K^{(\rho,\gamma)}=e\frac{f_\rho}{m_\rho}-{\rm Re}B_{\gamma\rho}(m_\rho^2)/m_\rho^2$. 
%The constants $f_\rho$ and $f_\omega$ correspond to the standard definitions of the leptonic decay constants 
%of the vector mesons (\ref{fV}). 

Let us summarise the parameters of our model. These are: 
\begin{itemize}
\item 
the Breit-Wigner masses of the vector mesons, that is $m_\rho$ and $m_\omega$, 
\item
 the decay constants $f_\rho$, $f_\omega$, 
\item 
the mixing parameter $b_{\rho\omega}$, 
\item
the couplings of the vector mesons $\rho$ and $\omega$ to two pions 
$g_{\rho\to\pi\pi}$, $g_{\omega\to\pi\pi}$, 
%$g_{\rho\to KK}$, $g_{\omega\to KK}$, $g_{\omega\to 3\pi}$, 
\item 
the width of the $\omega$ meson $\Gamma_\omega$,  
\item
and the $\pi\gamma$ couplings of the vector mesons 
$g_{\rho\to \pi\gamma}$ and $g_{\omega\to \pi\gamma}$. 
\end{itemize}
Values for these parameters can be found by the fit to the available form factor data. 
However, it turned out that the parameters $f_\omega$ and $\Gamma_{\omega}$ can not be 
well determined in this way from the reactions under discussion. The right place to extract these parameters from the experimental 
data is the reaction $e^+e^-\to \gamma^*\to 3\pi$, where the contribution of $\omega$ dominates. 
We leave a study of this reaction for a separate paper. Here 
we fix $f_\omega$ and $\Gamma_{\omega}$ to the PDG values, and leave only the remaining parameters from the
list above free in the fits.

%\begin{eqnarray}
%\label{su3}
%g_{\omega\to K^+K^-}=-g_{\omega\to K^0\bar K^0}
%=\sqrt{2}g_{\rho^0\to K^+K^-}=-\sqrt{2}g_{\rho^0\to K^0\bar K^0}
%=\frac{1}{\sqrt{2}}g_{\rho^0\to \pi^+\pi^-}. 
%\end{eqnarray}
Clearly, the inclusion of higher resonances with the same quantum numbers in the mixing scheme 
is straightforward. 

%*******************************************************************
\subsection{The form factors}
The calculation of the form factors is now straightforward: we must reconstruct the propagator matrix from its inverse, 
and then calculate the amplitude $\langle X|A_\mu|0\rangle$ from (\ref{5.12}) for the relevant final states.  
Finally we have to take into account that the amplitude of the e.m. current $\langle X(q)|J_\mu|0\rangle$ 
is related to the amplitude of the electromagnetic field as  
\begin{eqnarray}
\label{4.12}
\langle X(q)|J_\mu|0\rangle=\left(-g_{\mu\nu}q^2+q_\mu q_\nu\right)\langle X(q)|A^\nu|0\rangle.
\end{eqnarray}
For the pion form factor the final state is the $\pi\pi$ state, and for the $\pi\gamma$ form factor it is 
the $\pi\gamma$ state. Since we work to first order in the e.m. coupling, the set of the intermediate 
states is the same in both cases. In the model described above it includes the  $\pi\pi$, $3\pi$, and $KK$ 
intermediate states,  and we use  
(\ref{6.5}) and (\ref{6.5b}) for the vertex functions. 
To first order in the electromagnetic coupling the expressions for the pion elastic and 
the $\pi\gamma$ form factors obtained by the procedure described above may be written 
in a simple form (see Appendix \ref{app:C}) 
\begin{eqnarray}
\label{fpiel}
F_\pi(s)&=&1-s G_{V\to\pi\pi}^T\tilde\Delta(s) G_{\gamma\to V},
\\
\label{ffsmixedres}
F_{\gamma\pi}(s)&=&F_{\gamma\pi}(0)
-s G_{V\to\pi\gamma}^T \tilde\Delta(s) G_{\gamma\to V}, 
\end{eqnarray} 
with $F_{\gamma\pi}(0)$ given by (\ref{4.11}).
The propagator matrix $\tilde\Delta$ here is the inverse of the vector meson block of 
the matrix $\Delta^{-1}_T$ (\ref{6.7}) 
\begin{eqnarray}
\tilde\Delta^{-1}=
\left(\begin{array}{cc} 
-m_\rho^2+s+B_{\rho\rho}(s)                      &  s\,b_{\rho\omega}+B_{\rho\omega}(s) \\
s\,b_{\rho\omega}+B_{\rho\omega}(s)    & -m_\omega^2+s+B_{\omega\omega}(s)  
\end{array}\right)
\end{eqnarray} 
and 
\begin{eqnarray}
{G}_{V\to\pi\pi}=
\left(\begin{array}{c} 
\frac{1}{2}g_{\rho\to\pi\pi} \\
\frac{1}{2}g_{\omega\to\pi\pi}  
\end{array}\right),  
\qquad 
{G}_{V\to\pi\gamma}=
\left(\begin{array}{c} 
\frac{g_{\rho\to\pi\gamma}}{m_\rho} \\
\frac{g_{\omega\to\pi\gamma}}{m_\omega}  
\end{array}\right),   
\qquad 
{G}_{\gamma\to V}=
\left(\begin{array}{c} 
\frac{f_\rho}{m_\rho}+\frac{B_{\gamma\rho}}{e\,s}
\\
\frac{f_\omega}{m_\omega}+\frac{B_{\gamma\omega}}{e\,s}  
\end{array}\right).
\end{eqnarray} 
In the case of the charged-current form factor $F_\pi^+$ (\ref{weak}) 
describing the $\pi^-\to \pi^0$ transition the $\omega$ 
contribution is absent so we find from (\ref{f16}) of Appendix \ref{app:C}
\begin{eqnarray}
\label{piffweak0}
F_{\pi}^+(s)=
\frac{m_\rho^2-s+\frac{1}{2}g_{\rho\to\pi\pi} \frac{f_\rho s}{m_\rho} 
}{m_\rho^2-s-B_{\rho\rho}(s)}. 
\end{eqnarray}
Note that in the expressions for the pion electromagnetic and $\pi\gamma$ form factors 
(\ref{fpiel}) and (\ref{ffsmixedres}), the parameters $m_\rho$ and $f_\rho$ are those of the $\rho^0$ meson,
whereas in the formula for the charged-current form factor (\ref{piffweak0}) the parameters refer to the 
$\rho^-$ meson. 

The expression (\ref{piffweak0}) can be written in the "usual" vector meson dominance form (\ref{piff_c})
\begin{eqnarray}
\label{piffweak1}
F_{\pi}^+(s)=
\frac{\frac{1}{2}g_{\rho\to\pi\pi}f^{\rm eff}_\rho(s) m_\rho}{m_\rho^2-s-B_{\rho\rho}(s)}, 
\end{eqnarray}
in terms of an effective $s$-dependent $\gamma\rho$ coupling   
\begin{eqnarray}
\label{frhoeff}
f^{\rm eff}_\rho(s)=f_\rho\frac{s}{m_\rho^2}+\frac{2(m_\rho^2-s)}{g_{\rho\to\pi\pi}m_\rho}.   
\end{eqnarray} 
In this way we can make contact with the results of \cite{weise} where an $s$-dependent $\gamma\rho$ 
coupling is defined in an effective Lagrangian approach. 

We use the expressions (\ref{ffsmixedres}) and (\ref{piffweak0}) for the numerical analysis of the data for the pion electromagnetic,  
charge current, and $\pi\gamma$ transition form factors in the next Section.

%%\newpage
%************************************ fit1
\begin{table}%[bh]
\caption{\label{table:fit1}
Parameters of the resonances as found by the fit to the form factors 
in the region $\sqrt{s}< 0.9$ GeV within the $\gamma-\rho-\omega$ mixing scheme (Fit I). 
Higher resonances are not included. The PDG value for $f_\omega$ is used.  
%Charged and neutral $\rho$ mesons are treated separately: 
Fit to $F_\pi$ and $F_{\gamma\pi}$: $\chi^2/DOF=75/74$.  
Fit to the charged-current form factor $F^+_\pi$: $\chi^2/DOF=11/23$.  
The $\rho\omega$ mixing parameter has the value $b_{\rho\omega}=(3.5\pm 0.6)10^{-3}$.}
\centering
\vspace{.5cm}
\begin{tabular}{|c|l|c|c|c|}
\hline
Res. &  $m_V$, MeV &  $f_V$, MeV & 
$g_{V\to\pi\gamma}$ & 
$g_{V\to \pi\pi}$ \\
\hline
$\rho^-$   &  $775.5 \pm 0.4$     &  152.5$\pm$0.33               &            $-$   &  11.52$\pm$ 0.04  \\ 
$\rho^0$   &  $773.6 \pm 0.5$     &  154.1$\pm$0.67               &  0.60$\pm$ 0.06  &  11.43$\pm$ 0.04  \\ 
$\omega$   &  $782.42\pm 0.04$    &  $\underline{45.3\pm 0.9}$    &  1.79$\pm$ 0.09  &  $-0.27\pm$ 0.13  \\  
\hline
\end{tabular}
%************************************ fit2
\caption{\label{table:fit2}
Parameters of the resonances as found by the fit to the form factors 
in the region $\sqrt{s}< 0.9$ GeV (Fit II). 
The $\gamma-\rho-\omega$ mixing scheme with addition of the $\rho'=\rho(1450)$ is employed. 
The PDG values for $f_\omega$ and $m_{\rho'}$ are used.  
% Charged and neutral $\rho$ are treated separately. 
Fit to $F_\pi$ and $F_{\gamma\pi}$: $\chi^2/DOF=68/72$.  
Fit to $F^+_\pi$: $\chi^2/DOF=11/21$. 
The parameters of $\rho'$ cannot be determined by this fit. 
The $\rho\omega$ mixing parameter has the value $b_{\rho\omega}=(3.7\pm 0.6)10^{-3}$.}
\centering
\vspace{.5cm}
\begin{tabular}{|c|l|c|c|c|}
\hline
Res. &  $m_V$, MeV &  $f_V$, MeV & 
$g_{V\to\pi\gamma}$ & 
$g_{V\to \pi\pi}$ \\
\hline
$\rho^-$   &  $775.3 \pm 0.8$     &  152.4$\pm$0.4                &            $-$   &  11.50$\pm$ 0.05  \\ 
$\rho^0$   &  $773.8 \pm 0.6$     &  155.3$\pm$3.2                &  0.61$\pm$ 0.06  &  11.53$\pm$ 0.10  \\ 
$\omega$   &  $782.43\pm 0.05$     &  $\underline{45.3\pm 0.9}$   &  1.76$\pm$ 0.09  &  $-0.31\pm$ 0.10  \\  
\hline
$\rho'$    &  \underline{$1465 \pm 25$}       & $-$                  &  $-$    &  $-$  \\ 
\hline
\end{tabular}
\caption{\label{table:fit3}
Fit to the data for $\sqrt{s}\le 1.2$ GeV (Fit III), where 
$\rho$, $\omega$, $\rho'=\rho(1450)$, and $\rho''=\rho(1700)$ are taken into account. 
%The PDG value $f_\omega=45.3\pm 0.9$ is used.  
%$\rho^0$ and $\rho^-$ mesons are treated separately.   
Fit to $F_\pi$ and $F_{\gamma\pi}$: $\chi^2/DOF=72/89$.  
Fit to $F^+_\pi$: $\chi^2/DOF=13/27$.  
The extracted $\rho$ and $\omega$ couplings and masses are very stable with respect to 
inclusion/exclusion of $\rho''$. The couplings of $\rho'$ 
and $\rho''$ correlate very strongly and cannot be reliably determined by this fit.  
%We use the PDG values $f_\omega=45.3\pm 0.9$ and $m_\rho'=1465\pm 25$ MeV. 
%The parameters $g_{\rho'\to\pi\gamma}$ cannot be determined by the fit. \\
The $\rho\omega$ mixing parameter has the value $b_{\rho\omega}=(3.5\pm 0.5)10^{-3}$.}
\begin{tabular}{|c|l|c|c|c|}
\hline
Res. &  $m_V$, MeV &  $f_V$, MeV & 
$g_{V\to\pi\gamma}$ & 
$g_{V\to \pi\pi}$ \\
\hline
$\rho^-$   &  $775.3 \pm 0.5$     &  151.5$\pm$1.5              &  $-$             &  11.50$\pm$ 0.05  \\ 
$\rho^0$   &  $773.7 \pm 0.4$     &  155.4$\pm$1.7              &  0.65$\pm$0.05   &  11.51$\pm$ 0.07 \\ 
$\omega$   &  $782.43\pm 0.05$    &  $\underline{45.3\pm 0.9}$  &  1.73$\pm$0.08   &  $-0.35\pm 0.10$ \\  
\hline
$\rho'$    & $\underline{1465\pm 25}$ &  $-$         &  $-$              &   $-$  \\ 
$\rho''$   & $\underline{1700\pm 20}$ &  $-$         &  $-$              &   $-$  \\ 
\hline
\end{tabular}
\end{table}

\newpage
\section{Numerical results}
We analyse the recent data on the pion electromagnetic form factor 
$F_\pi$ (\ref{2.2}), the charged-current form factor $F^+_{\pi}$ (\ref{weak}), and 
$F_{\gamma\pi}$ form factor (\ref{2.13}) in the region $\sqrt{s}=0\div 1.2$ GeV
using the formulae (\ref{fpiel}), (\ref{ffsmixedres}), and (\ref{piffweak0}), respectively. 
We take into account that the pion electromagnetic form factor $F_\pi$ and the $\pi\gamma$ transition 
form factor $F_{\gamma\pi}$ contain contributions of the neutral $\rho^0$ and $\omega$ resonances, 
whereas the charged-current form factor $F^+_\pi$ contains the contribution of the $\rho^-$ meson. 
Since we consider the isospin-violating $\rho^0-\omega$ mixing effects, we do not 
assume the parameters of the charged and the neutral $\rho$ mesons to be equal to each other.  
We therefore fit the data for the $F_\pi$ and $F_{\gamma\pi}$ form factors and extract in this way the 
$\omega$ and $\rho^0$ parameters. We use the recent SND data \cite{data2} for the form factor 
$F_{\gamma\pi}$  and the recent update \cite{data1} of the 
CMD-2 data \cite{data0} for $F_\pi$. We also include the available data on the phase 
of the electromagnetic form factor \cite{phase}.  

We separately fit the form factor $F^+_\pi$ and extract the $\rho^-$ parameters from 
the CLEO data \cite{cleo}. 

We perform different fitting procedures explained below. The fitted values of our parameters are 
given in Tables \ref{table:fit1},\ref{table:fit2}, and \ref{table:fit3}. 

The first fitting procedure (Fit I) includes the $\rho^-$, $\rho^0$, 
and $\omega$ contributions, and neglects effects of higher resonances.
We work in the region $\sqrt{s}\le 0.9$ GeV where this approximation 
is checked to be self-consistent. 
The $\chi^2/DOF$ for this fit is $75/74$ for $F_{\pi}$ and $F_{\gamma\pi}$. It is 
$11/23$ for $F^+_\pi$. The resulting values 
of the parameters are given in Table \ref{table:fit1}.

In the second step (Fit II), we study the stability of the extracted $\rho$ and $\omega$ parameters 
with respect to the inclusion of higher resonances. This might serve as a probe of the systematic errors. 
We still stay in the region $\sqrt{s}\le 0.9$ GeV, but include in addition to $\rho$ and $\omega$ 
also the $\rho'=\rho(1450)$ resonance, assuming its contribution to have the form (\ref{piff_c})
with the corresponding mass and coupling constant of $\rho'$. We fix the mass of this 
resonance according to PDG \cite{pdg} and neglect possible $\rho-\rho'$ mixing effects. 
The parameters obtained by this procedure are given in Table \ref{table:fit2}. The quality of the 
fit to $F_{\pi}$ and $F_{\gamma\pi}$ definitely improves ($68/72$), whereas for $F^+_{\pi}$ no 
improvement is seen. It is worth noting that the extracted parameters of the $\rho$ and $\omega$ 
turn out to be very stable with respect to the inclusion/exclusion of $\rho'$. 

\begin{figure}% [hb]
\begin{center}
\begin{tabular}{c}
\mbox{\epsfig{file=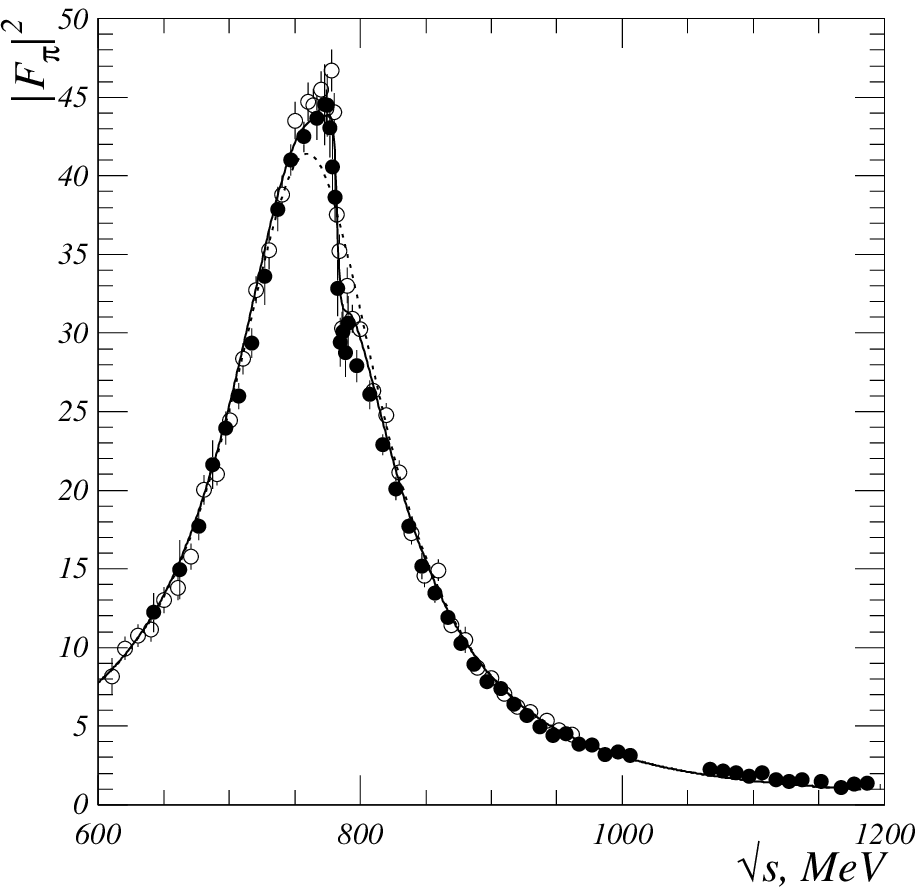,height=10cm}}  \\
\mbox{\epsfig{file=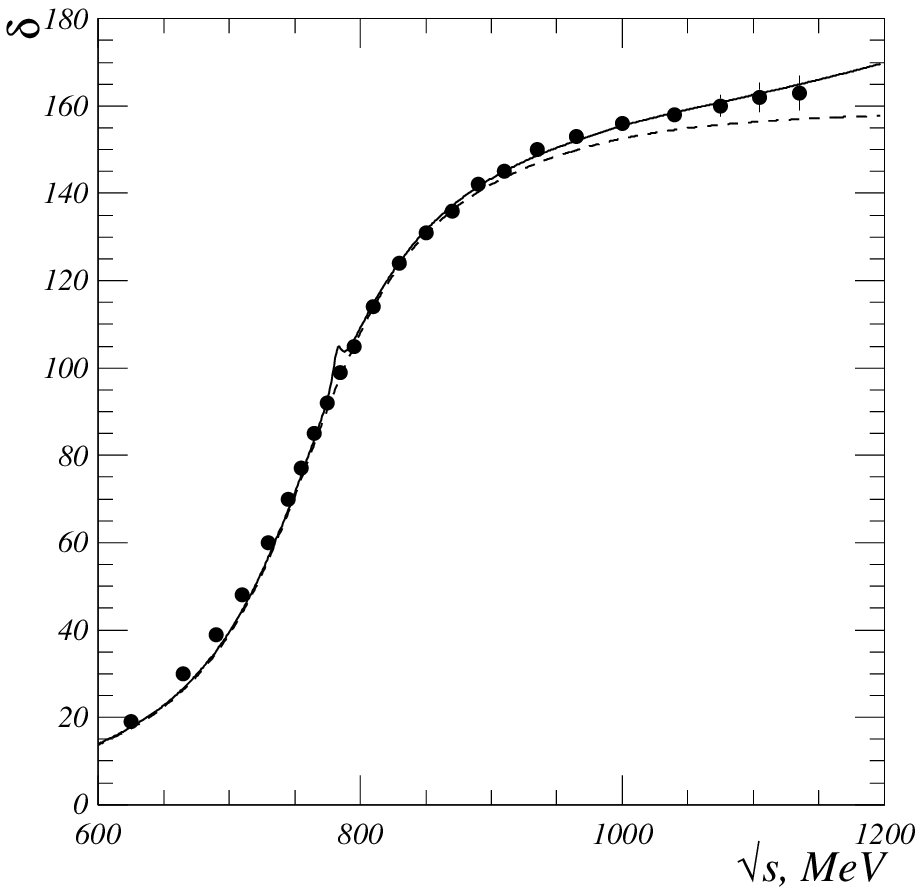,height=10cm}}
\end{tabular}
\caption{\label{fig:piemff}
The pion electromagnetic form factor $F_{\pi}(s)$: 
(a) Modulus squared: data from \cite{data1} (empty) used in the fit. 
Data from \cite{datapifftimelike} (full) which was not used for the fit is shown for comparison. 
(b) Phase: data from \cite{phase}. 
Solid lines - the full form factor as obtained by Fit III, 
dashed lines - the $\rho$-contribution in Eq. (\ref{fpiel}). }
\end{center}
\end{figure}

\begin{figure} %[hb]
\begin{center}
\mbox{\epsfig{file=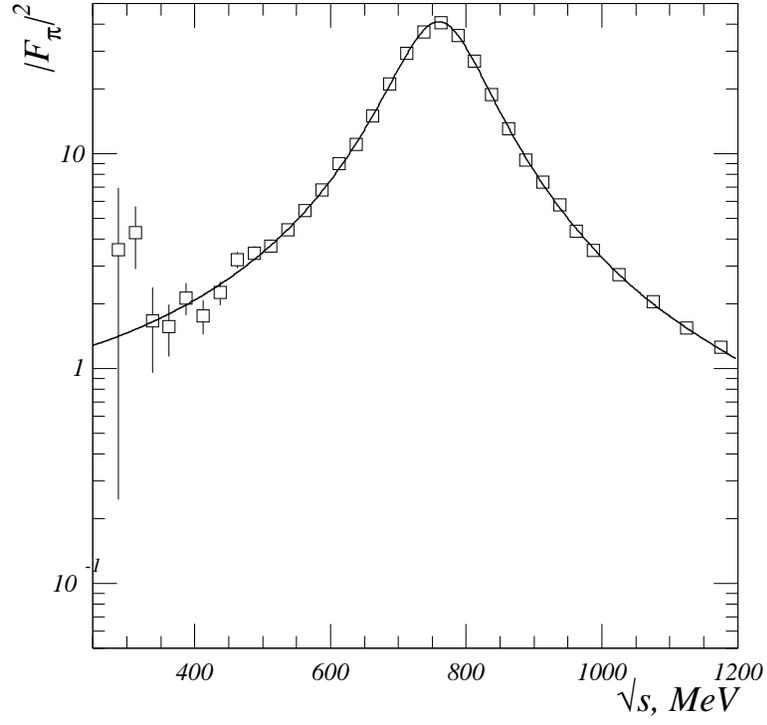,height=10cm}} 
\caption{\label{fig:piweakff}
The weak transition $\pi^-\to\pi^0$ form factor, Fit III. Data from \cite{cleo}.}
\end{center}
\end{figure}

\begin{figure} %[hb]
\begin{center}
\mbox{\epsfig{file=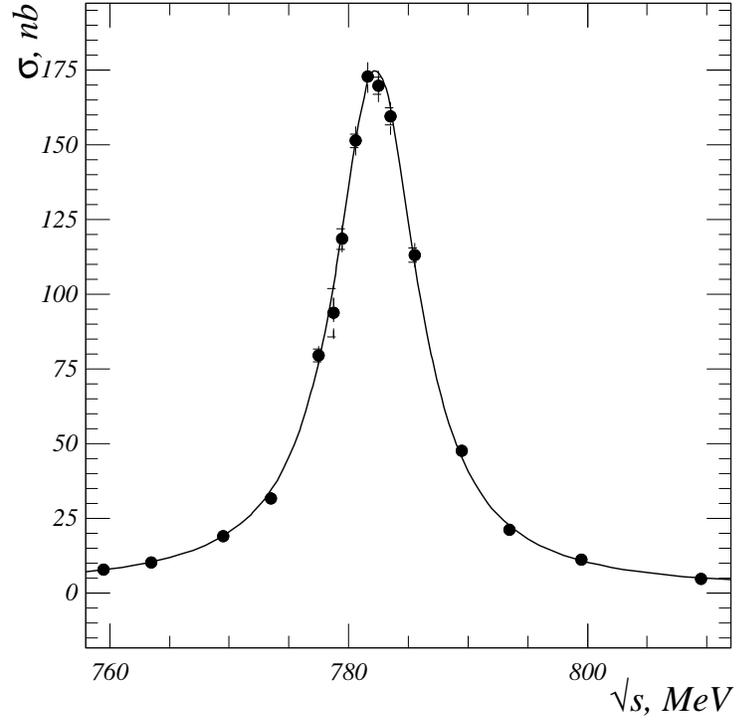,height=10cm}} 
\caption{\label{fig:cross}
The cross-section $\sigma_{e^+e^-\to\pi\gamma}(Q^2)$, Eq. (\ref{cross-section}) 
calculated with $F_{\gamma\pi}$ from Eq. (\ref{ffsmixedres}). Data from \cite{data2}.}
\end{center}
\end{figure}

In the third step, we extend our analysis to the region $\sqrt{s}\le 1.2$ GeV (Fit III). 
We fit the form factors taking into account the three resonances $\rho$, $\omega$, and $\rho'$
($\chi^2/DOF=72/89$ for $F_{\pi}$ and $F_{\gamma\pi}$; $\chi^2/DOF=13/27$ for $F^+_{\pi}$). 
The results of this fit are given in Table \ref{table:fit3}.  
We then also include in addition to them the $\rho''=\rho(1700)$. 
The coupling constants of $\rho'$ and $\rho''$ turn out to be strongly correlated with each 
other and therefore cannot be extracted from the data under consideration. For a
reliable extraction of these parameters one should go to higher values of $s$. Important for our 
analysis is that the masses and couplings of the $\rho$ and $\omega$ mesons are remarkably stable 
with respect to inclusion/exclusion of $\rho''$ and very well compatible with the numbers obtained in
Fits I and II. The form factors calculated with the parameters from 
Table \ref{table:fit3} are shown in Figs \ref{fig:piemff}, \ref{fig:piweakff}, and \ref{fig:cross} 
as solid lines. 

The very satisfactory description of the data speaks in favour of the reliability of our assumptions 
of the dominance of the $\pi\pi$, $KK$, and $3\pi$ intermediate states and on the 
negligible $s$-dependence of the vertex functions. 

Note that the masses of the charged and the neutral $\rho$ mesons are different 
as obtained by our fits exposing an isospin violation in $\rho$ mesons which is extensively 
discussed in the literature (see \cite{jeger} and refs therein). 

However we would like to point out that assuming the parameters 
of the charged and the neutral $\rho$ mesons to be equal to each other also leads to a very 
good description of the data with $\chi^2/DOF$ below 1 for all fitting procedures I-III
(Fit I: $\chi^2/DOF=94/100$, Fit II: $\chi^2/DOF=83/98$, Fit III: $\chi^2/DOF=86/121$). 
Therefore, strictly speaking, the data analysed by us here does not require 
the masses and couplings of the charged and neutral $\rho$ mesons to be different. 
Still, we do not think it reasonable to take into account isospin-violating $\rho-\omega$ mixing 
effects and assume the absence of such effects in the charged and neutral $\rho$ mesons. 
We therefore do not discuss in detail the fitting procedures in which the parameters of the charged 
and neutral mesons are assumed to be equal. 

\begin{figure} %[hb]
\begin{center}
\mbox{\epsfig{file=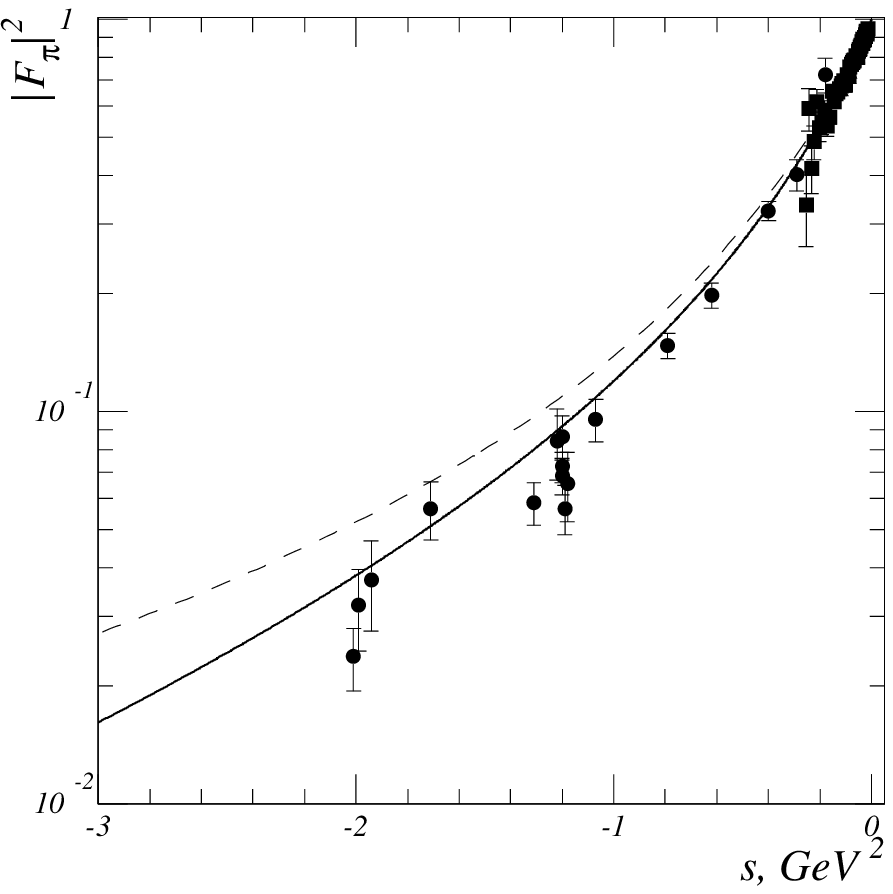,height=10cm}} 
\caption{\label{fig:piemffspacelike}
The pion electromagnetic form factor: $|F_{\pi}(s)|^2$ for $s<0$.  
Data from \cite{datapiffspacelike}. Solid line: the full form factor 
from fit III, 
%Eq. (\ref{ffsmixedres}), 
dashed line: $F_{\pi}=1/(1-s/m_\rho^2)$.}
\end{center}
\end{figure}

Fig. \ref{fig:piemffspacelike} shows the pion form factor at small spacelike momentum 
transfers, which was not included in the fit. Still one can see a good agreement with the data 
even down to $s=-2$ GeV${}^2$. The improvement of the description of $F_\pi$ compared to the 
naive VMD Ansatz (\ref{piff_a}) shown as dashed line is obvious. Similar results were obtained in 
\cite{weise}. 

Next we study the low energy expansion of the pion 
electromagnetic form factor near $s=0$ 
\begin{eqnarray}
\label{paramff}
F_{\pi}(s)=1+\frac{1}{6}\langle r^2 \rangle^\pi_V+c^\pi_V s^2 +O(s^3). 
\end{eqnarray}
Here $\langle r^2 \rangle^\pi_V$ is the squared charge radius of the pion. This quantity and $c^\pi_V$ 
are of great interest in the framework of ChPT for fixing certain parameters, see for instance 
\cite{pich1,8}. In Table \ref{table:coeff} we compare our results with those of \cite{pich1,8}  
and the naive VMD. We find full consistency with ChPT within the errors which for our results are only 
statistical ones. Of course a study of systematic errors should also be done but this is beyond the 
scope of this paper.

\begin{center}
\begin{table}%[b]
\caption{\label{table:coeff}The pion form factor at small momentum transfers. 
Coefficients of the expansion (\protect\ref{paramff}) from fit III are given. The errors given 
in our results are only statistical emerging from errors in masses and couplings.}  
\begin{tabular}{|r|l|l|}
\hline
                                &   $\langle r^2 \rangle^\pi_V$, GeV${}^{-2}$  &  $c^\pi_V$, GeV${}^{-4}$   \\
\hline			     
Our result                      &         11.41$\pm$0.05   & 3.83$\pm$0.02   \\
ChPT to order $O(p^6)$ \cite{8} &         11.22$\pm$0.41   & 3.85$\pm$0.6   \\
Results from \cite{pich1}        &         11.04$\pm$0.3    & 3.79$\pm$0.04   \\ 
Naive VMD Eq. (\ref{piff_a})    &         10.16            &    2.8          \\
\hline
\end{tabular}  
\end{table}
\end{center}
Finally, let us discuss the $\rho$ and $\omega$ masses and the $\rho$ width. 
The value of the $\rho$-meson Breit-Wigner mass defined in (\ref{6.2}) $m_\rho=774.3\pm 0.4$ MeV 
agrees with the value obtained recently from the weak pion form factor \cite{pich2}, 
and is sizeably higher than the value $m_\rho=771.1\pm0.9$ MeV quoted by PDG.  

The Breit-Wigner width of the $\rho$ meson is defined according to the relation \cite{pich2} 
\begin{eqnarray}
1/\Gamma^{BW}_\rho=m_\rho \frac{d\delta_\pi(s)}{ds}|_{s=m_\rho^2}, 
\end{eqnarray}
where $\delta_\pi(s)$ is the phase of the pion form factor $F_\pi^+(s)$. Numerically, 
we obtain from fit III for the charged $\rho$ 
$$
\Gamma^{\rm BW}_{\rho}=149.85\pm 0.4\;{\rm MeV}. 
$$
Next we turn to the pole masses and widths of the $\rho^-$, $\rho^0$ and $\omega$.  
For the charged $\rho$-meson the location of the pole in the second Riemann sheet of the $s$-plane 
is found by solving the equation 
\begin{eqnarray}
m_\rho^2-s-B_{\rho\rho}(s)=0. 
\end{eqnarray}
The corresponding solution, $s_{\rm pole}$, can be used in two different ways 
to define pole masses and pole widths. Either we choose to set
\begin{eqnarray}
\label{pole1}
s_{\rm pole}={M'}_\rho^2-i\Gamma^{'\rm pole}_\rho {M'}_\rho,
\end{eqnarray}
or 
\begin{eqnarray}
\label{pole2}
s_{\rm pole}=(M_\rho-i\Gamma^{\rm pole}_\rho/2)^2. 
\end{eqnarray}
The values obtained with the definition (\ref{pole2}) are given in Table \ref{table:numerical}.

The pole masses of the $\rho^0$ and the $\omega$ meson are affected by the $\rho\omega$ mixing 
effects and are obtained from the equation, see (\ref{f6}) 
\begin{equation}
\left\{m_\rho^2-s-B_{\rho\rho}(s)\right\}\left\{m_\omega^2-s-B_{\omega\omega}(s)\right\}-
\left\{s\,b_{\rho\omega}+B_{\rho\omega}(s)\right\}^2=0. 
\end{equation}
The corresponding values of the pole masses of the $\rho^0$ and $\omega$ and widths are also given in 
Table \ref{table:numerical}. 
\begin{center}
% pole masses are calculated with rho1.nb
\begin{table}%[thb]
\caption{\label{table:numerical}Pole masses and widths of $\rho$ and $\omega$.} 
\begin{tabular}{|r|r|r|r|r|}
\hline
 & $\rho^{+}$ & $\rho^{0}$  &  $\omega$\\
 \hline
$M^{\rm pole}$, MeV        &  756.7$\pm$0.4   &   755.8$\pm$0.4   &     782.44$\pm$0.05   \\
\hline
$\Gamma^{\rm pole}$, MeV   &  144.7$\pm$0.4   &   143.8$\pm$0.4   &      8.38$\pm$0.05   \\
\hline
\end{tabular}  
\end{table}
\end{center}
%%%\newpage
%________________________________________________________________
%**********************************************************************************
%**********************************************************************************
%**********************************************************************************
%**********************************************************************************
%%\newpage
\section{Conclusions}
We have discussed a general approach to the description of vector mesons and their mixing 
using dispersion relations. This approach allows us to represent various observables for the vector 
mesons as products of the vector-meson propagator matrix and the reduced amplitudes (vertex functions).  
The unitarity relation gives the anti-hermitian part of the propagator matrix in terms of the 
relevant reduced amplitudes. 

The merit of this approach lies in the possibility to separate different types of singularities 
in different quantities: the propagator matrix of the vector-meson fields contains the 
resonance poles 
which lead to "fast" variations of the form factors in the resonance region, 
whereas the reduced amplitudes are free from these singularities and are therefore slowly varying 
functions in the resonance region. 
The description is fully general at this stage and contains no approximations. To go further 
we need some dynamical inputs for the reduced amplitudes. 

We then formulate our model for the form factors based on the following assumptions:  
\begin{itemize}
\item[(i)] 
we take into consideration  the resonances $\rho$, $\omega$, and in a rough way also the $\rho(1450)$ 
and neglect higher vector mesons;
\item[(ii)] 
we take into account the $\pi\pi$, $K\bar K$ and effectively also $3\pi$ intermediate states,  
and neglect contributions of multi-meson states in the unitarity relations;  
\item[(iii)]
we assume the scalar coupling factors in the reduced amplitudes to be constant in the region of the 
momentum transfer $\sqrt{s}=0 - 1.2$ GeV. 
\end{itemize}
On the basis of these assumptions we perform a combined analysis of the recent data for 
several reactions: $e^+e^-\to \pi^+\pi^-$, $\tau^-\to \pi^-\pi^0\nu_\tau$, 
and $e^+e^-\to \pi^0\gamma$ in the region of $\sqrt{s}=0-1.2$ GeV. 
All the analysed data is well described in our approach, allowing for an extraction 
of the resonance parameters, such as the Breit-Wigner masses and effective coupling constants.  
Our main numerical results are given in Tables \ref{table:fit1}-\ref{table:fit3}. 

These results are obtained by the fitting procedures, which allow for different masses and couplings 
of the charged and the neutral $\rho$ mesons. Therefore, we have fitted separately the charged 
current form factor $F^+_\pi$ and the neutral current form factors $F_\pi$ and $F_{\gamma\pi}$. 
Still we would like to point out that assuming the equality of the parameters of the charged and 
the neutral $\rho$ mesons and fitting all three form factors $F^+_\pi$, $F_\pi$, and $F_{\gamma\pi}$
simultaneously also leads to a good description of the data with $\chi^2/DOF$ below unity. 
However since the $\rho-\omega$ mixing and the isospin violation in $\rho$ mesons 
are both effects of the same nature, we do not think it reasonable to include only the first of these
effects. We therefore do not discuss in detail the hypothesis that the masses and couplings of charged
and neutral $\rho$ mesons are equal. 

The small errors of our results are statistical errors only indicating a good description 
of the data by our form factor formulae. Still our form factors are based on certain model assumptions,  
therefore a systematic error should be added. 

One way to estimate the systematic error is to vary the fit range, to choose different 
parametrizations for the reduced amplitudes, and to study the corresponding variations 
of the fitted quantities. Partially, we did this by choosing various ranges of $s$ and including 
higher resonances in our fits. We observed a good stability of the extracted resonance parameters 
and therefore do not expect the systematic error to be large. 
A more detailed study of the systematic errors is left for future work. 

Further topics to be studied in the future are the inclusion of the $\phi$ meson in the mixing scheme 
and the implications of our form factor formulae for the theoretical values of the $g-2$ factor of the muon and
the fine structure constant at the $Z$ mass, $\alpha_s(m_Z)$.  

To summarise, we have obtained a very good description of the form factors in a model based on the 
unitarity and dispersion relations. Chiral perturbation theory constraints were checked to be respected. 
We have introduced reduced amlitudes or vertex functions which describe the coupling of the photon $\gamma$ 
and of the vector meson $\rho$ and $\omega$ fields to two pseudoscalar mesons and to $\pi\gamma$. 
These vertex functions contain invariant coupling functions which in principle depend on the momentum 
transfer $\sqrt{s}$. Perhaps the most interesting result of our study is that these invariant coupling 
functions are really coupling \underline{constants} for $0\le\sqrt{s}\le 1.0$ GeV. Thus these couplings 
are "frozen" below the GeV scale. This may perhaps be related to the expectation mentioned frequently in the 
literature \cite{carlo} that in QCD the coupling parameter $\alpha_s(\mu^2)$ is "frozen" for $\mu^2\lesssim$ 1 GeV${}^2$.  

%%\newpage
\section{Acknowledgments}
We are grateful to V.~V.~Anisovich, N.~Brambilla, E.~De~Rafael, A.~Donnachie, C.~Ewerz, M.~Jamin, F.~Nagel, 
and B.~Stech for useful discussions and to J. Urheim for correspondence on the CLEO data. 
We also thank M.~Jamin for a careful reading of the manuscript. 
This work was supported by the Alexander 
von Humboldt-Stiftung and Bundesministerium f\"ur Bildung und Forschung project 05 HT 1VHA/0. 

%***************************************************************************************
%%\newpage

%*********************APPENDICES***********************

\appendix
\newpage

\section{\label{app:A}Gauge invariance and properties of the matrices $M^2$ and $K$}
Here we study the conditions to be imposed on the propagator matrix (\ref{3.1}) arising from the   
requirements to have a massless photon, the correct charge normalisation, and no 
strong-interaction long-range force. This leads to the following constraints at $q^2\to 0$: 
\begin{eqnarray}
\label{a1}
{\Delta}_T^{(1,1)}(q^2)&=&\frac{1}{q^2}\left(1+O(q^2)\right), \\
\label{a2}
{\Delta}_T^{(i,j)}(q^2)&=&O(1), \quad {\rm for}\;(i,j)\ne (1,1).
\end{eqnarray} 
These conditions require the matrices $M^2$ and $K$ to satisfy 
\begin{eqnarray}
\label{a3}
M^2_{1j}&=&0, \quad {\rm for}\;j=1,\ldots,n, \\
\label{a4}
K_{11}&=&1.
\end{eqnarray} 
To show this, we note first that by a transformation of the type (\ref{5.23}), but involving only 
$V_{\mu}^{(j)}$ with $j>1$, we can always achieve 
\begin{eqnarray}
\label{a5}
M^2_{1j}&=&0, \quad {\rm for}\;j>2.
\end{eqnarray} 
Then the matrix $M^2$ takes the form 
\begin{eqnarray}
\label{a6}
M^2=
\left(\begin{array}{c|ccc} 
M^2_{11} & M^2_{12} & 0   \quad \ldots& 0    \\
\hline 
M^2_{12} &  &  &   \\
0        &  &  &   \\
.        &  & M^{2'} &  \\
0        &  &  &  \\
\end{array}\right), 
\end{eqnarray} 
where 
\begin{eqnarray}
\label{a7}
M^{2'}=
\left(\begin{array}{ccc} 
M^2_{22} & \ldots&  M^2_{2n}    \\
.  & \ldots & .  \\
M^2_{n2} & \ldots&  M^2_{nn}    
\end{array}\right). 
\end{eqnarray} 
For $n>2$ we define 
\begin{eqnarray}
\label{a8}
M^{2''}=
\left(\begin{array}{ccc} 
M^2_{33} & \ldots&  M^2_{3n}    \\
.  & \ldots & .  \\
M^2_{n3} & \ldots&  M^2_{nn}    
\end{array}\right), 
\end{eqnarray} 
and for $n=2$ we set $M^{2''}=1$.

In the following we assume that $M^{2'}$ is a positive-definite matrix. 
This means, for instance, that all fileds $V^{(j)}_{\mu}$, $j=2,\ldots,n$ 
must be independent. 

From (\ref{5.21}) and (\ref{a6}) we find now for $q^2\to 0$
\begin{eqnarray}
\label{a9}
{\rm det\,}{\Delta}^{-1}_T(q^2)&=&(-1)^n {\rm det\,}M^2+O(q^2)
\nonumber\\
&=&(-1)^n\left[ M^2_{11} {\rm det\,} M^{2'}-\left(M^2_{12}\right)^2 
{\rm det\,} M^{2''}\right]+O(q^2), 
\end{eqnarray} 

\begin{eqnarray}
\label{a10}
{\Delta}^{(1,1)}_T(q^2)&=&
\frac{ {\rm det\,}(-M^{2'})+O(q^2) }{ {\rm det\,}{\Delta}^{-1}_T(q^2) }
\nonumber\\
&=&
\frac{-{\rm det\,} M^{2'}+O(q^2) }{{\rm det\,} M^{2}+O(q^2)  },
\end{eqnarray} 

\begin{eqnarray}
\label{a11}
{\Delta}^{(1,2)}_T(q^2)&=&
\frac{ (-1)^n M^2_{12}{\rm det\,} M^{2''}+O(q^2) }{ {\rm det\,}{\Delta}^{-1}_T(q^2) }
\nonumber\\
&=&
\frac{M^2_{12}{\rm det\,}M^{2''}+O(q^2)}{{\rm det\,} M^{2}+O(q^2) },
\end{eqnarray} 
We see now from (\ref{a10}) that in order to fulfil (\ref{a1}) we must have 
${\rm det\,} M^{2}=0$. Then (\ref{a2}) for $(i,j)=(1,2)$ and (\ref{a11}) require $M_{12}^2=0$. 
Combining this result with (\ref{a9}) gives $M_{11}^2=0$. Recall that we assumed 
$M^{2'}>0$ which implies also $M^{2''}>0$. Thus, we have already demonstrated (\ref{a3}). 
Inserting now (\ref{a3}) into (\ref{5.21}) gives 
\begin{eqnarray}
\label{a12}
{\rm det\,}{\Delta}^{-1}_T(q^2)&=&(-1)^{n-1} q^2 K_{11}{\rm det\,}M^{2'}+O\left((q^2)^2\right).
\end{eqnarray} 
With (\ref{a10}) this leads to 
\begin{eqnarray}
\label{a13}
{\Delta}^{(1,1)}_T(q^2)&=&
\frac{ (-1)^{n-1}{\rm det\,} M^{2'}+O(q^2) }
{ (-1)^{n-1}q^2 K_{11}{\rm det\,} M^{2'}+O\left((q^2)^2\right)}
\nonumber\\
&=&
\frac{1}{q^2 K_{11}}\left(1+O(q^2) \right). 
\end{eqnarray} 
To fulfil (\ref{a1}) we must, therefore, have $K_{11}=1$ which proves (\ref{a4}).

%*********** APPENDIX B ************

\section{\label{app:B} The inverse propagator matrix for the $\gamma-\rho-\omega$ system}

Given (\ref{6.4}), the inverse propagator (\ref{5.21}) for the $\gamma-\rho-\omega$ system 
can be written as 
\begin{eqnarray}
\label{b1}
\Delta^{-1}_{T}(s)=-\left(\begin{array}{ccc} 
0 &  0       & 0 \\
0 & m_\rho^2 & 0 \\
0 & 0        & m_\omega^2 
\end{array}\right)+s K+s^2\frac{1}{\pi}
\int\limits_{4m_\pi^2}^{\infty}ds'\frac{D_T(s')}{s'^2(s'-s-i\epsilon )}. 
\end{eqnarray}
Here we set $q^2=s$ and $D_T(s')$ is obtained using Assumptions 1 and 2 of Section IV and 
Eq. (\ref{5.17}). 

It is convenient to split the constant matrix $K$ into two parts 
\begin{eqnarray}
\label{b2}
K=K^{a}+K^{b}, 
\end{eqnarray}
to be specified later, and to define the matrix function 
\begin{eqnarray}
\label{b3}
B(s)=s K^a+s^2\frac{1}{\pi}
\int\limits_{4m_\pi^2}^{\infty}ds'\frac{D_T(s')}{s'^2(s'-s-i\epsilon )}. 
\end{eqnarray}
We thus have 
\begin{eqnarray}
\label{b4}
\frac{1}{2i}\left\{B(s)-B^\dagger(s)\right\}=D_T(s). 
\end{eqnarray}
The matrix $K^a$ will be chosen such that it cancels the real part of 
the dispersive contribution to $B(s)$ in Eq (\ref{b3}) at some 
specific values of $s$. Then splitting the matrix $K$ according to (\ref{b2}) 
has an unambiguous physical meaning. 

We choose $K^a_{\gamma\gamma}=0$.
With this, the condition (\ref{44b}) requires $K^b_{\gamma\gamma}=1$. 

The diagonal elements $K^a_{VV}$, $V=\rho,\omega$, are chosen such that they cancel the real 
part of the dispersive contribution in (\ref{b3}) at $s=m_V^2$ for the vector meson $V$. 
That is, we require
\begin{eqnarray}
\label{b5}
{\rm Re}\, B_{\rho\rho}(m_\rho^2)=0, \qquad {\rm Re}\, B_{\omega\omega}(m_\omega^2)=0. 
\end{eqnarray}
The normalisations for the $\rho$ and $\omega$ fields are fixed by requiring (\ref{6.4}).  
Given (\ref{b5}), this requirement leads to 
\begin{eqnarray}
\label{b6}
K^b_{\rho\rho}=K^b_{\omega\omega}=1. 
\end{eqnarray}

The non-diagonal matrix elements 
$(K^a)_{\gamma V}$ are chosen to cancel the real part of the dispersive 
contribution to $B_{\gamma V}$ at the point $s=m_V^2$. 
That is, we require 
\begin{eqnarray}
\label{b8}
{\rm Re}\, B_{\gamma\rho}(m_\rho^2)=0, \qquad {\rm Re}\, B_{\gamma\omega}(m_\omega^2)=0. 
\end{eqnarray}
Now we set 
\begin{eqnarray}
\label{b9}
(K^b)_{\gamma V}=e\frac{f_V}{m_V}, 
\end{eqnarray}
where the real parameters $f_V$ correspond to our precise definitions of the leptonic decay
constants of the vector mesons $V=\rho,\omega$. 
 
%For the non-diagonal matrix elements describing the mixing of the vector mesons 
%(in particular the $\rho-\omega$ mixing) 
%there is some ambiguity in choosing the reference point at which the subtraction 
%is done. One of the possibiliites is to choose this point at $q^2=m_\rho^2$.   
%Then the matrix element 

Finally, the matrix element 
$(K^a)_{\rho\omega}$ is chosen such that it cancels the real part of the dispersive 
contribution to $B_{\rho\omega}$ in (\ref{b3}) at the point $s=m_\rho^2$. 
Thus we require 
\begin{eqnarray}
\label{b10}
{\rm Re}\,B_{\rho\omega}(m_\rho^2)=0.
\end{eqnarray}
Then the element 
\begin{eqnarray}
\label{b11}
(K^b)_{\rho\omega}=b_{\rho\omega}
\end{eqnarray}
defines our $\rho-\omega$ mixing parameter. Clearly the value of the 
mixing parameter depends on the choice of the subtraction point in (\ref{b10}), but 
physics, of course, does not.

% ********* Meson loop diagrams *********
%
%The analysis of coupling constants shows that for the mixing in the system of light 
%vector mesons the diagrams of Fig. \ref{fig:mixing2} give the main contribution and 
%have to be taken into account. These diagrams determine the functions $B_{ij}$ in the 
%inverse propagator matrix (\ref{6.7}). 
%
%The imaginary part of the loop diagram with light pseudoscalars in the intermediate state 
%can be calculated unambiguosly from the Feynman expression, and the real part can be 
%reconstructed by dispersion relations which need two subtractions.  %
%
%We fix one subtraction constant requiring that both diagonal and non-diagonal diagrams 
%with the intermediate light pseudoscalars vanish at $s=0$. 
%
%The second subtraction constants in the diagonal functions $B_{ii}$ is fixed by 
%requiring ${\rm Re}\, B_{ii}(m_i^2)=0$. 
%
%The non-diagonal function $B_{\rho\omega}$ appears in combination with 
%the direct $\rho\omega$ mixing amplitude $b_{\rho\omega}$. Therefore, the 
%second subtraction constant in the non-diagonal function can be absorbed in $b_{\rho\omega}$ 
%to be found by the fit \cite{mnp}. 
%

Let us now discuss in more detail each of the functions $B_{ij}(s)$ of (\ref{b3}).
%
%  FIGURE
%\begin{figure}%[b]
%\begin{center}
%\mbox{\epsfig{file=ro-mixing.eps,width=17cm}}
%\caption{\label{fig:mixing2}
%Diagrams to be taken into account for the description of the light vector meson mixing: 
%(a). the dominant diagrams for $B_{\rho\rho}$	  ($\pi\pi$, $K^+K^-$ and $K^0\bar K^0$ intermediate states) 
%(b). the dominant diagrams for $B_{\omega\omega}$ ($3\pi$, $K^+K^-$ and $K^0\bar K^0$ intermediate states), 
%(c). the direct $\rho\omega$ mixing amplitude and the dominant diagram for $B_{\rho\omega}$ ($\pi\pi$ intermediate states). 
%The $K^0\bar K^0$ and $K^+K^-$ intermediate state contributions to 
%$B_{\rho\omega}$ cancel each other. 
%A comment on $B_{\rho\phi}$: 
%it does not receive contributions from the intermediate two and three pion
%states because of the vanishing $g_{\phi\to \pi\pi}$ 
%and $g_{\rho\to 3\pi}$ coupling constants. Contributions of the $K^0\bar K^0$ and $K^+K^-$ 
%intermediate state cancel each other. Therefore we set $B_{\rho\phi}=0$
%}
%\end{center}
%\end{figure} 
%
%************************************************************************************
%%%\newpage
\subsubsection{$B_{\rho\rho}$}
The function ${\rm Im}\,B_{\rho\rho}$ receives contributions from the 
$\pi^+\pi^-$, $K^+K^-$  and $K^0\bar K^0$ intermediate states. 
The two-pion contribution to ${\rm Im}\,B_{\rho\rho}$ reads  
\begin{eqnarray}
\label{80}
{\rm Im}\; B_{\rho\rho}(s)|_{\pi\pi}&=&g_{\rho\to\pi\pi}^2 {\rm Im}\; B_{\pi\pi}(s)
\end{eqnarray}
where 
\begin{eqnarray}
\label{is}
{\rm Im}\; B_{\pi\pi}(s)&=&I(s,m_\pi^2), 
\nonumber\\
I(s,m^2)&=&\frac{1}{192\pi}s\left(1-\frac{4m^2}{s}\right)^{3/2}\theta(s-4m^2).  
\end{eqnarray}
We have to take into account also $K^+K^-$ and $K^0\bar K^0$ 
intermediate states which give contributions similar to (\ref{80}), (\ref{is}) with 
$\pi$ replaced by $K$. 
The coupling constant $g_{\rho\to KK}$ cannot be measured directly so we 
use the flavour-SU(3) relations (\ref{su3})
\begin{eqnarray}
\label{gsu3}
2g_{\rho\to KK} =g_{\rho\to \pi\pi}.  
\end{eqnarray}
Then we find 
\begin{eqnarray}
\label{relationim1}
{\rm Im}\; B_{\rho\rho}(s) &=&g_{\rho\to \pi\pi}^2 
\left[{
{\rm Im} \;B_{\pi\pi} +
\frac{1}{4}\left(
{\rm Im}\;B_{K^+K^-}+{\rm Im}\;B_{K^0\bar K^0}\right)
}\right]
%\nonumber\\&=&
=g_{\rho\to \pi\pi}^2 \left[{
{\rm Im}\; B_{\pi\pi} +
\frac{1}{2}{\rm Im}\;B_{KK}}\right].    
\end{eqnarray}
As explained above, see (\ref{b4}), we require 
\begin{eqnarray}
{\rm Re}\; B_{\rho\rho}(m_\rho^2)=0. 
\end{eqnarray}
Putting everything together we find from (\ref{b3})
%and fix the second subtraction constant by setting  
%\begin{eqnarray}
%B_{\rho\rho}(s=0)=0. 
%\end{eqnarray}
%As the result we find 
\begin{widetext}
\begin{eqnarray}
\label{83}
B_{\rho\rho}(s)=g_{\rho\to \pi\pi}^2\;s
\left[
{R(s,m_\pi^2)-R(m^2_{\rho},m_\pi^2)}
+\frac{R(s,m_K^2)-R(m^2_{\rho},m_K^2)}{2}
\right]
+ig_{\rho\to \pi\pi}^2\;\left[
I(s,m_\pi^2)+\frac{I(s,m_K^2)}{2} \right]. 
\end{eqnarray}
Here $I(s,m^2)$ is defined in (\ref{is}),  and 
\begin{eqnarray}
\label{r}
R(s,m^2)&=&\frac{s}{192\pi^2}\;{\rm V.P.}\;\int_{4m^2}^\infty \frac{ds'}{(s'-s)s'}\left(1-\frac{4m^2}{s'}\right)^{3/2}
\\
\nonumber
&=&\left\{\begin{array}{lll} 
\frac{1}{96\pi^2}
\left(
\frac{1}{3}+\xi^2+\frac{\xi^3}{2}\log\left(\frac{1-\xi}{1+\xi}\right)
\right),  
& \qquad \xi=\sqrt{1-\frac{4m^2}{s}}, 
& \quad {\rm for}\;\; s>4m^2, \\
\frac{1}{96\pi^2}\left(
\frac{1}{3}-\xi^2+\xi^3\,{\rm arctan}\left(\frac{1}{\xi}\right)
\right), 
& \qquad \xi=\sqrt{\frac{4m^2}{s}-1}, 
& \quad {\rm for}\;\; 0<s<4m^2, \\  
\frac{1}{96\pi^2}
\left(
\frac{1}{3}+\xi^2+\frac{\xi^3}{2}\log\left(\frac{\xi-1}{\xi+1}\right)
\right),  
& \qquad \xi=\sqrt{1-\frac{4m^2}{s}}, 
& \quad {\rm for}\;\; s<0, \\
\end{array}\right. 
\end{eqnarray}
where V.P. means the principle value. 
\end{widetext}
%************************************************************************************
\subsubsection{$B_{\omega\omega}$}
When considering the $B_{\omega\omega}$, the intermediate $K^+K^-$, 
$K^0\bar K^0$, and $3\pi$ states should be taken into account. 

First, recall that the coupling constant $g_{\omega\to 3\pi}$ is 
much smaller than the coupling constant $g_{\omega\to KK}$. This becomes clear 
by using the SU(3) relation $g_{\omega\to KK}=\frac{1}{2}g_{\rho\to \pi\pi}$.  
Therefore, the contribution of the $3\pi$ intermediate states can be safely neglected
in the {\it real} part of $B_{\omega\omega}$.  

However, the decay $\omega\to K\bar K$ is forbidden kinematically. Due to this, 
the $\omega$ width emerges mainly due to the $3\pi$ intermediate states and is 
small because $g_{\omega\to 3\pi}$ is small. 
Therefore, the $3\pi$ states are crucial for the calculation of the {\it imaginary} 
part of $B_{\omega\omega}$ below the $K\bar K$ threshold. Still, the $s$-dependence of 
${\rm Im}\;B_{\omega\omega}$ can in practice be neglected, 
and we may therefore use for it a constant width approximation 
$\Gamma_\omega={\rm const}$. Taking into account that $B_{\omega\omega}=0$ for $s=0$ (\ref{6.8}),  
and making use of (\ref{b5}), we can write with sufficient accuracy for our purposes 
\begin{eqnarray}
B_{\omega\omega}(s)=
g_{\omega\to KK}^2\;s
\left[
{R(s,m_K^2)-R(m^2_{\omega},m_K^2)}
\right]
+i \Gamma_\omega {s}/{m_\omega}. 
\end{eqnarray}
%************************************************************************************
\subsubsection{$B_{\gamma\rho}$}
The function ${\rm Im}\,B_{\gamma\rho}$ receives 
contributions from the $\pi^+\pi^-$ and $K^+K^-$ intermediate states. 
Taking into account the SU(3) relations (\ref{su3}) between the coupling constants we find
\begin{eqnarray}
{\rm Im}\; B_{\gamma\rho}&=&
2\,e\,g_{\rho\to\pi\pi}\left(
{\rm Im}\; B_{\pi\pi} +
\frac{1}{2}{\rm Im}\;B_{K^+K^-}\right)
%\nonumber\\&=&
=2\,e\,g_{\rho\to\pi\pi} \left(
{\rm Im} \;B_{\pi\pi} +\frac{1}{2}{\rm Im}\;B_{KK}\right),  
\end{eqnarray}
and hence from (\ref{83})
\begin{eqnarray}
{\rm Im}\; B_{\gamma\rho}(s)=\frac{2\,e}{g_{\rho\to\pi\pi}}\;{\rm Im}\; B_{\rho\rho}(s).  
\end{eqnarray}
With the conditions (\ref{b4}) and (\ref{b8}) we get also 
\begin{eqnarray}
\label{b21}
B_{\gamma\rho}(s)=\frac{2\,e}{g_{\rho\to\pi\pi}}B_{\rho\rho}(s).  
\end{eqnarray}
%*************************************************************************
\subsubsection{$B_{\gamma\omega}$}
The function ${\rm Im}\;B_{\gamma\omega}$ receives 
contributions from the $K^+K^-$ intermediate states:  
\begin{eqnarray}
{\rm Im}\; B_{\gamma\omega}&=&
2eg_{\omega\to KK}{\rm Im}\;B_{K^+K^-}.  
\end{eqnarray}
Requiring (\ref{b8}) we come to the expression  
\begin{eqnarray}
B_{\gamma\omega}(s)=2eg_{\omega\to KK}\;s
\left[
R(s,m_K^2)-R(m^2_{\omega},m_K^2)\right]
+2ieg_{\omega\to KK}\;I(s,m_K^2).  
\end{eqnarray}
%*************************************************************************
\subsubsection{$B_{\rho\omega}$}
The imaginary part of $B_{\rho\omega}$ is given by 
the $\pi\pi$ intermediate states so we have 
\begin{eqnarray}
{\rm Im}\; B_{\rho\omega}(s)&=&g_{\rho\to\pi\pi}g_{\omega\to\pi\pi} {\rm Im}\;B_{\pi\pi}(s).  
\end{eqnarray}
Making use of (\ref{b10}), $B_{\rho\omega}$ takes the form  
\begin{eqnarray}
\label{brhoomega}
B_{\rho\omega}(s)=g_{\rho\to \pi\pi}g_{\omega\to\pi\pi}
\left[
{s\,R(s,m_\pi^2)-s\,R(m^2_{\rho},m_\pi^2)+iI(s,m_\pi^2)}\right].  
\end{eqnarray}
This completes our discussion of the individual functions $B_{ij}(s)$. The resulting inverse
propagator matrix is given in (\ref{6.7}). For comparison with (\ref{5.21}) we also list the
resulting matrix elements of $K$ 
\begin{eqnarray}
K_{\gamma\gamma}&=&1,  
\nonumber\\
K_{\rho\rho}&=&1-g^2_{\rho\to\pi\pi}
\left[R(m_\rho^2,m_\pi^2)+\frac{1}{2}R(m_\rho^2,m_K^2)\right],
\nonumber\\
K_{\omega\omega}&=&1-g^2_{\omega\to KK}R(m_\omega^2,m_K^2),
\end{eqnarray}
\begin{eqnarray}
K_{\gamma\rho}=K_{\gamma\rho}&=&e\frac{f_\rho}{m_\rho}-2eg_{\rho\to\pi\pi}
\left[R(m_\rho^2,m_\pi^2)+\frac{1}{2}R(m_\rho^2,m_K^2)\right],
\nonumber\\
K_{\gamma\omega}=K_{\gamma\omega}&=&e\frac{f_\omega}{m_\omega}-2eg_{\omega\to KK}R(m_\omega^2,m_K^2),  
\nonumber\\
K_{\rho\omega}=K_{\rho\omega}&=&b_{\rho\omega}-
g_{\omega\to \pi\pi}g_{\rho\to \pi\pi}R(m_\rho^2,m_\pi^2), 
\nonumber 
\end{eqnarray} 
with the functions $R$ given in (\ref{r}). 

Using the values for the masses and coupling constants from Fit III, we find the following central
values for $K_{ij}$:
$K_{\rho\rho}=1.017$, $K_{\omega\omega}=0.994$, $K_{\rho\omega}=2.6\cdot 10^{-3}$, 
$K_{\gamma\rho}=0.2 \,e$, $K_{\gamma\omega}=0.06\, e$, 
$e=\sqrt{4\pi\alpha_{\rm e.m.}}\simeq 1/3$. 
Clearly, the deviation of the matrix $K$ from the $3\times 3$ unit matrix is small, 
only at the percent level. 
%\newpage 

Finally, we discuss the relation of our expressions to the original Gounaris-Sakurai expression for
the form factor. The formula given in (11) of \cite{gs} can be written as 
\begin{eqnarray} 
\label{b27}
F_\pi^{\rm GS}(s)=\frac{m_\rho^2+d\,m_\rho\Gamma_\rho^{\rm GS}}{m_\rho^2-s-B^{\rm GS}_{\rho\rho}(s)}, 
\end{eqnarray} 
where 
\begin{eqnarray}
\label{b28}
B^{\rm GS}_{\rho\rho}(s)=-\Gamma_\rho^{\rm GS}\left(\frac{m_\rho^2}{k_\rho^3}\right)
\left\{
k^2[h(s)-h(m_\rho^2)]+k_\rho^2 h'(m_\rho^2)(m_\rho^2-s)
\right\}
+im_\rho\Gamma_\rho^{\rm GS}\left(k/k_\rho\right)^3m_\rho/\sqrt{s}, 
\end{eqnarray}
\begin{eqnarray}
\label{b29}
k=\left\{\begin{array}{lll} 
&(\frac{1}{4}s-m_\pi^2)^{1/2} \quad &{\rm for}\;\;\quad  s\ge 4m_\pi^2,\\ 
&i(m_\pi^2-\frac{1}{4}s)^{1/2} \quad &{\rm for}\;\;\quad  0\le s < 4m_\pi^2,
\end{array}\right.
\end{eqnarray}
\begin{eqnarray}
\label{b30}
d&=&\frac{3}{\pi}\frac{m_\pi^2}{k_\rho^2}\log\left(\frac{m_\rho+2k_\rho}{2m_\pi}\right)+
\frac{m_\rho}{2\pi k_\rho}-\frac{m_\pi^2 m_\rho}{\pi k^3_\rho},
\end{eqnarray}
\begin{eqnarray}
\label{b31}
h(s)&=&\frac{2}{\pi}\frac{k}{\sqrt{s}}\log\left(\frac{\sqrt{s}+2k}{2m_\pi}\right). 
\end{eqnarray}
The only free parameters in the GS formula are $m_\rho$ and $\Gamma_\rho^{\rm GS}$.
It is easy to see that the following relations hold 
\begin{eqnarray}
\nonumber
&{\rm Re}\,B_{\rho\rho}^{\rm GS}(m_\rho^2)=0,\\
\nonumber
&\frac{d}{ds}{\rm Re}\,B_{\rho\rho}^{\rm GS}(s)|_{s=m_\rho^2}=0,\\
\label{b32}
&B_{\rho\rho}^{\rm GS}(0)=-d\,m_\rho\, \Gamma_\rho^{\rm GS}.
\end{eqnarray}
This proves (\ref{piff_b}) and (\ref{piff_gsb}).
Comparing (\ref{b28}) to (\ref{is}) and (\ref{r}) and setting 
\begin{eqnarray}
\label{b33}
\Gamma_\rho^{\rm GS}=g^2_{\rho\to\pi\pi}\frac{1}{24\pi}\frac{k_\rho^3}{m_\rho^2}
\left(1-\frac{B_{\rho\rho}^{\rm GS}(0)}{m_\rho^2} \right)
\end{eqnarray}
we obtain 
\begin{eqnarray}
\label{b34}
B_{\rho\rho}^{\rm GS}(s)&=&\left(1-\frac{B_{\rho\rho}^{\rm GS}(0)}{m_\rho^2} \right)
\tilde B_{\rho\rho}(s) -\frac{B_{\rho\rho}^{\rm GS}(0)}{m_\rho^2}(s-m_\rho^2), \\
\label{b35}
F^{\rm GS}_\pi(s)&=&\frac{m_\rho^2}{m_\rho^2-s-\tilde B_{\rho\rho}(s)}. 
\end{eqnarray}
Here $\tilde B_{\rho\rho}(s)$ is defined as our $B_{\rho\rho}(s)$ in (\ref{83}) but 
omitting the $K\bar K$ contributions.  

To compare the GS with our expression it is best to choose $F_\pi^+(s)$ (\ref{piffweak0}) 
since no $\omega$ contributions are included in (\ref{b35}). Clearly, also the $K\bar K$ 
contributions are not included in (\ref{b35}), but one could easily do so replacing 
$\tilde B_{\rho\rho}(s)$ by the full $B_{\rho\rho}(s)$. The remaining difference between 
(\ref{b35}) and (\ref{piffweak0})-(\ref{frhoeff}) is in the $s$-dependence of the effective 
$\rho-\gamma$ coupling. The GS formula would correspond to the relation 
\begin{eqnarray}
\label{b36}
\frac{1}{2}g_{\rho\to\pi\pi}\frac{f_\rho}{m_\rho}=1.
\end{eqnarray}
Using the $\rho^-$ parameters from our Fit III gives, however, 
\begin{eqnarray}
\label{b37}
\frac{1}{2}g_{\rho\to\pi\pi}\frac{f_\rho}{m_\rho}=1.12\pm 0.05.
\end{eqnarray}

\section{\label{app:C} Propagator matrix and form factors} 
In this appendix we give the derivation of the expressions (\ref{fpiel}), 
(\ref{ffsmixedres}), and (\ref{piffweak0}) for the form factors. 
We start with writing the inverse propagator matrix (\ref{6.7}) as follows
\begin{eqnarray}
\label{f1}
\Delta_T^{-1}(s)=
\left(\begin{array}{c|c} 
s        &  \quad 0   \\
\hline 
0        & \tilde\Delta^{-1}(s) 
\end{array}\right)
+e\,s 
\left(\begin{array}{c|c} 
0        &  G^T_{\gamma\to V}  \\
\hline 
G_{\gamma\to V}        & 0 
\end{array}\right), 
\end{eqnarray} 
where 
\begin{eqnarray}
\label{f2}
\tilde\Delta^{-1}(s)=
\left(\begin{array}{cc} 
-m_\rho^2+s+B_{\rho\rho}(s)                      &  s\,b_{\rho\omega}+B_{\rho\omega}(s) \\
s\,b_{\rho\omega}+B_{\rho\omega}(s)    & -m_\omega^2+s+B_{\omega\omega}(s)  
\end{array}\right)
\end{eqnarray} 
and 
\begin{eqnarray}
\label{f3}
{G}_{\gamma\to V}=
\left(\begin{array}{c} 
\frac{f_\rho}{m_\rho}+\frac{B_{\gamma\rho}}{e\,s}
\\
\frac{f_\omega}{m_\omega}+\frac{B_{\gamma\omega}}{e\,s}  
\end{array}\right),  
\end{eqnarray} 
The tranverse propagator to order $e$ then reads 
\begin{eqnarray}
\label{f4}
\Delta_T(s)&=&\left\{
1+e
\left(\begin{array}{c|c} 
0        &       G^T_{\gamma\to V}\\
\hline
s\tilde\Delta(s)G_{\gamma\to V}  &  0
\end{array}\right)
\right\}^{-1}
\left(\begin{array}{c|c} 
1/s        &       0\\
\hline
0  &  \tilde\Delta(s)
\end{array}\right)
\nonumber\\
&=&
\left(\begin{array}{c|c} 
1/s        &   -eG^T_{\gamma\to V}\tilde\Delta(s)    \\
\hline
-e \tilde\Delta(s)G_{\gamma\to V}   &  \tilde\Delta(s)
\end{array}\right)+O(e^2), 
\end{eqnarray} 
\begin{eqnarray}
\label{f5}
\tilde\Delta(s)=({\rm det}\;\tilde\Delta^{-1}(s))^{-1} 
\left(\begin{array}{cc} 
-m_\omega^2+s+B_{\omega\omega}(s)               & -s\,b_{\rho\omega}-B_{\rho\omega}(s) \\
-s\,b_{\rho\omega}-B_{\rho\omega}(s)    & -m_\rho^2+s+B_{\rho\rho}(s)  
\end{array}\right),
\end{eqnarray} 
\begin{eqnarray}
\label{f6}
{\rm det}\;\tilde\Delta^{-1}(s)=
(-m_\omega^2+s+B_{\omega\omega}(s))(-m_\rho^2+s+B_{\rho\rho}(s))-(s\,b_{\rho\omega}+B_{\rho\omega}(s))^2.   
\end{eqnarray}

The pole masses and the pole widths of the $\rho^0$ and $\omega$ are obtained as solutions of 
\begin{eqnarray}
\label{f6a}
{\rm det}\,\Delta_T^{-1}(s)=s\cdot {\rm det}\,\tilde\Delta^{-1}(s)=0. 
\end{eqnarray}
To obtain the expression for the electromagnetic form factor we use now 
(\ref{2.2}), (\ref{4.12}), (\ref{5.12}), and (\ref{5.13}).
This gives with $q=p+p'$, $s=(p+p')^2=q^2$
\begin{eqnarray}
\label{f7}
e(p'-p)_\mu F_\pi(q^2)
&=&
\langle \pi^+(p')\pi^-(p)|J_\mu(0)|0\rangle
\nonumber\\
&=& 
\langle \pi^+(p')\pi^-(p)|A^\nu(0)|0\rangle \left(-g_{\mu\nu}q^2+q_{\mu}q_{\nu}\right)
\nonumber\\
&=& 
\langle \pi^+(p')\pi^-(p)||V^{(i)\lambda}(0)||0\rangle 
\Delta^{(i,1)}_{\lambda\nu}(q) 
\left(-\delta^{\nu}{}_\mu q^2+q^{\nu}q_{\mu}\right)
\nonumber\\
&=& 
\langle \pi^+(p')\pi^-(p)||V^{(i)\lambda}(0)||0\rangle 
\Delta^{(i,1)}_{T}(s)\left(g_{\lambda\mu}q^2-q_{\lambda}q_{\mu}\right)
\nonumber\\
&=& 
\langle \pi^+(p')\pi^-(p)||V^{(i)}_{T\mu}(0)||0\rangle\, s\,\Delta^{(i,1)}_{T}(s)
\end{eqnarray}
Inserting here (\ref{6.5}), (\ref{6.6a}) and (\ref{f4}) leads to 
\begin{eqnarray}
\label{f8}
F_\pi(s)&=&1-s G_{V\to\pi\pi}^T\tilde\Delta(s) G_{\gamma\to V},
\end{eqnarray}
where 
\begin{eqnarray}
\label{f9}
{G}_{V\to\pi\pi}=
\left(\begin{array}{c} 
\frac{1}{2}g_{\rho\to\pi\pi} \\
\frac{1}{2}g_{\omega\to\pi\pi}  
\end{array}\right),  
\end{eqnarray} 
We can further simplify the expression for $F_{\pi}$ in (\ref{f8}) taking into account 
that $g_{\omega\to\pi\pi}$ is much smaller than $g_{\rho\to\pi\pi}$, and that 
the $\rho\omega$ mixing parameter is also small as comes out from the fit. 
From (\ref{f6}) and (\ref{brhoomega}) we get then 
\begin{eqnarray}
\label{f12}
{\rm det}\;\tilde\Delta^{-1}(s)=
(-m_\omega^2+s+B_{\omega\omega}(s))(-m_\rho^2+s+B_{\rho\rho}(s))
+O(g^2_{\omega\to\pi\pi},b^2_{\rho\omega},g_{\omega\to\pi\pi}b_{\rho\omega}) 
\end{eqnarray}
For $F_\pi(s)$ of (\ref{f8}) this gives 
\begin{eqnarray}
\label{f13}
F_\pi(s)&=&1+
\frac{1}{2}{g_{\rho\to\pi\pi}}
\frac{
\frac{f_\rho}{m_\rho}s+\frac{1}{e}B_{\gamma\rho}(s)
}{m_\rho^2-s-B_{\rho\rho}(s)}
\nonumber\\
&+&
\frac{1}{2}{g_{\rho\to\pi\pi}}
\frac{
\left(\frac{f_\omega}{m_\omega}s+\frac{1}{e}B_{\gamma\omega}(s)\right)
\left(s b_{\rho\omega}+B_{\rho\omega}(s)\right)
}{\left(m_\rho^2-s-B_{\rho\rho}(s)\right)\left(m_\omega^2-s-B_{\omega\omega}(s)\right)}
+
\frac{1}{2}{g_{\omega\to\pi\pi}}
\frac{
\frac{f_\omega}{m_\omega}s+\frac{1}{e}B_{\gamma\omega}(s)
}{m_\omega^2-s-B_{\omega\omega}(s)}
\nonumber\\
&+&O(g^2_{\omega\to\pi\pi},b^2_{\rho\omega},g_{\omega\to\pi\pi}b_{\rho\omega}).  
\end{eqnarray} 
In a similar way we get for the $\gamma\pi$ transition form factor (\ref{2.13}) 
\begin{eqnarray}
\label{f10}
F_{\gamma\pi}(s)&=&F_{\gamma\pi}(0)
-s G_{\gamma\to V}^T \tilde\Delta(s) G_{V\to\pi\gamma}, 
\end{eqnarray} 
with $F_{\gamma\pi}(0)$ given by the anomaly (\ref{fpigamma0}) and 
\begin{eqnarray}
\label{f11}
{G}_{V\to\pi\gamma}=
\left(\begin{array}{c} 
\frac{g_{\rho\to\pi\gamma}}{m_\rho} \\
\frac{g_{\omega\to\pi\gamma}}{m_\omega}  
\end{array}\right).
\end{eqnarray} 
Expanding in $g_{\omega\to\pi\pi}$ and $b_{\rho\omega}$ gives 
\begin{eqnarray}
\label{f14}
F_{\gamma\pi}(s)&=&F_{\gamma\pi}(0)+
\frac{g_{\rho\to\gamma\pi}}{m_\rho}
\frac{
\frac{f_\rho}{m_\rho}s+\frac{1}{e}B_{\gamma\rho}(s)
}{m_\rho^2-s-B_{\rho\rho}(s)}
+
\frac{g_{\omega\to\gamma\pi}}{m_\omega}
\frac{
\frac{f_\omega}{m_\omega}s+\frac{1}{e}B_{\gamma\omega}(s)
}{m_\omega^2-s-B_{\omega\omega}(s)}
\nonumber\\
&+&
\frac{s b_{\rho\omega}+B_{\rho\omega}(s)}
{\left(m_\rho^2-s-B_{\rho\rho}(s)\right)\left(m_\omega^2-s-B_{\omega\omega}(s)\right)}
\left[
\frac{g_{\rho\to\gamma\pi}}{m_\rho}\left(\frac{f_\omega}{m_\omega}s+
\frac{1}{e}B_{\gamma\omega}(s)\right) 
+
\frac{g_{\omega\to\gamma\pi}}{m_\omega}\left(\frac{f_\rho}{m_\rho}s+
\frac{1}{e}B_{\gamma\rho}(s)\right)
\right]
\nonumber\\
&+&O(g^2_{\omega\to\pi\pi},b^2_{\rho\omega},g_{\omega\to\pi\pi}b_{\rho\omega}).  
\end{eqnarray} 
The weak form factor (\ref{weak}) 
$F_\pi^+$ is obtained from (\ref{f13}) setting $g_{\omega\to\pi\pi}=0$ and $b_{\rho\omega}=0$. 
This gives
\begin{eqnarray}
\label{f16} 
F_\pi^+(s)=\frac{1}{m_\rho^2-s-B_{\rho\rho}(s)}
\left(
m_\rho^2-s+\frac{1}{2}g_{\rho\to\pi\pi}\frac{f_\rho s}{m_\rho}-B_{\rho\rho}(s)
+\frac{g_{\rho\to\pi\pi}}{2e}B_{\gamma\rho}(s)\right). 
\end{eqnarray} 
Using now (\ref{b21}) leads to (\ref{piffweak0}).
\end{document}